%% file: main.tex
\newif\ifAnon\Anonfalse
\newif\ifDraft\Draftfalse
\newif\ifSubmission\Submissiontrue

\documentclass[letterpaper,twocolumn,10pt]{article}
\usepackage[hyphens]{url}
\PassOptionsToPackage{bookmarks,hidelinks}{hyperref}
\usepackage{hyperref}
\hypersetup{breaklinks=true}
\usepackage{usenix2020}
\usepackage{cispa_defaults}
\usepackage{todonotes}
\usepackage{balance}

\usepackage[firstpage]{draftwatermark}
\SetWatermarkText{\hspace*{6in}\raisebox{7.5in}{\includegraphics{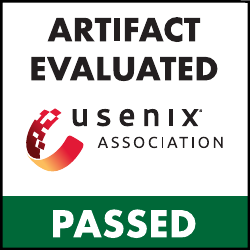}}}
\SetWatermarkAngle{0}

\renewcommand{\paragraph}[1]{\vspace{0.0cm}\noindent\textbf{#1}\ }
\renewcommand{\subparagraph}[1]{\vspace{0.0cm}\textit{#1}\ }

\begin{document}
\title{\Large \bf Osiris: Automated Discovery of  \\
Microarchitectural Side Channels}
\date{}

\ifAnon

\else
\author{
  {\rm Daniel Weber, Ahmad Ibrahim, Hamed Nemati, Michael Schwarz, Christian Rossow}\\
  CISPA Helmholtz Center for Information Security
}
\fi

\maketitle

\newcommand{\ToolName}{Osiris\xspace}
\newcommand{\KASLRBreak}{FlushConflict\xspace}
\newcommand{\MovReload}{Stream+Reload\xspace}

\newcommand{\sZero}{${S}_0$\xspace}
\newcommand{\sOne}{${S}_1$\xspace}
\newcommand{\sTwo}{${S}_2$\xspace}

\newcommand{\styleSequence}[1]     {\ensuremath{\mathsf{#1}}}
\newcommand{\styleScript}[1] {\ensuremath{\texttt{#1}}}
\newcommand{\styleComponent}[1]     {\ensuremath{\mathit{#1}}}
\newcommand{\styleArchitecture}[1] {{\fontfamily{DejaVuSansMono}\selectfont{#1}}}
\newcommand{\styleParameter}[1]  {\ensuremath{\mathsf{#1}}}
\newcommand{\styleSignal}[1]     {\ensuremath{\mathsf{#1}}}
\newcommand{\styleCommand}[1]     {\ensuremath{\mathit{#1}}}

\newcommand{\ReSeq}              {\styleSequence{Seq_{reset}}\xspace}
\newcommand{\TrSeq}              {\styleSequence{Seq_{trigger}}\xspace}
\newcommand{\MeSeq}              {\styleSequence{Seq_{measure}}\xspace}

\newcommand{\reasonMI}{\raisebox{-0.2em}{\includegraphics[height=1em]{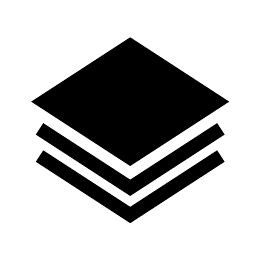}}}
\newcommand{\reasonHT}{\raisebox{-0.2em}{\includegraphics[height=1em]{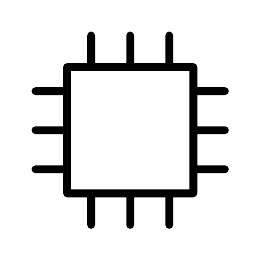}}}
\newcommand{\reasonOP}{\raisebox{-0.2em}{\includegraphics[height=1em]{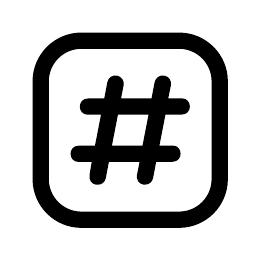}}}

\begin{abstract} %
In the last years, a series of side channels have been discovered on CPUs. 
These side channels have been used in powerful attacks, \eg on cryptographic implementations, or as building blocks in transient-execution attacks such as Spectre or Meltdown. 
However, in many cases, discovering side channels is still a tedious manual process. 

In this paper, we present \ToolName, a fuzzing-based framework to automatically discover microarchitectural side channels.
Based on a machine-readable specification of a CPU's ISA, \ToolName generates instruction-sequence triples and automatically tests whether they form a timing-based side channel. 
Furthermore, \ToolName evaluates their usability as a side channel in transient-execution attacks, \ie as the microarchitectural encoding for attacks like Spectre. 
In total, we discover four novel timing-based side channels on Intel and AMD CPUs. 
Based on these side channels, we demonstrate exploitation in three case studies.
We show that our microarchitectural KASLR break using non-temporal loads, \KASLRBreak, even works on the new Intel Ice Lake and Comet Lake microarchitectures. 
We present a cross-core cross-VM covert channel that is not relying on the memory subsystem and transmits up to \SI{1}{\kilo\bit/\second}. 
We demonstrate this channel on the AWS cloud, showing that it is stealthy and noise resistant.
Finally, we demonstrate \MovReload, a covert channel for transient-execution attacks that, on average, allows leaking 7.83 bytes within a transient window, improving state-of-the-art attacks that only leak up to 3 bytes. 
\end{abstract}

\section{Introduction} \label{sec:intro}
Since first described by Kocher~\cite{Kocher1996} in 1996, side channels have kept challenging the security guarantees of modern systems. 
Side channels targeted mostly cryptographic implementations in the beginning~\cite{Kocher1996,Osvik2006,Aciicmez2008,Gullasch2011}. 
By now, they have also been shown to be powerful attacks to spy on user behavior~\cite{Gruss2015Template,Monaco2018,Schwarz2018KeyDrown}. 
Moreover, in transient-execution attacks, such as Meltdown~\cite{Lipp2018meltdown} or Spectre~\cite{Kocher2019}, side channels are vital.

Side channels often arise from abstraction and optimization~\cite{Schwarz2019Thesis}. 
For example, due to the internal complexity of modern CPUs, the actual implementation, \ie the microarchitecture, is abstracted into the documented architecture. 
This abstraction also enables CPU vendors to introduce transparent optimizations in the microarchitecture without requiring changes in the architecture. 
However, these optimizations regularly introduce new side channels that attackers can exploit~\cite{Percival2005,Osvik2006,Acicmez2007new,Bhattacharya2012,Shin2018Prefetch,Sullivan2018minefields,Schwarz2019STL,Lipp2020takeaway}.

Although new side channels are commonly found, discovering a side channel typically requires manual effort and a deep understanding of the underlying microarchitecture. 
Moreover, with multiple thousand variants of instructions available on the x86 architecture alone~\cite{Abel2019uops}, the number of possible side effects that can occur when combining instructions is too large to test manually. 
Hence, manually identified side channels represent only a subset of the side channels of a CPU. 

Indeed, automatically finding CPU-based side channels is challenging.
Side channels consist of a carefully-chosen interplay of multiple orthogonal instructions that are syntactically far apart from each other.
Typically, they require instructions that change an inner CPU state and others reading (leaking) this inner state.
In addition, many side channels rely on specific instructions to reset the internal state to a known one.
For example, the popular Flush+Reload side channel~\cite{Yarom2014Flush} flushes cache lines to reset the state, fills a secret-dependent cache line, and uses another cache access to leak the new state.
Identifying such an interplay \emph{automatically} is notoriously hard, fueled by thousands of CPU instructions, their possible combinations, and the lack of mechanisms to verify the existence of potential side-channel candidates.

Automation attempts, therefore, have focused on particular types of side channels so far.
With Covert Shotgun and ABSynthe, Fogh~\cite{Fogh2016shotgun} and Gras~\etal\cite{Gras2020}, respectively, automated the discovery of contention-based side channels. 
Their tools identified several side effects of instructions when run simultaneously on the two logical cores, \ie hyperthreads, of a physical CPU core. 
However, their approach does not generalize beyond contention-based side channels. 
Moghimi~\etal\cite{Moghimi2020medusa} considered the sub-field of microarchitectural data-sampling (MDS) attacks. 
Their tool, Transynther, combines and mutates building blocks of existing MDS attacks to find new attack \emph{variants}. 
However, they do not try to find new classes of side channels, and only focus on cache-based covert channels.

In this paper, we present a generic approach to automatically detect timing-based side channels that do not rely on contention. 
We introduce a notation for side channels that allows representing side channels as triples of instruction sequences:
one that resets the inner CPU state (\emph{reset sequence}), one that triggers a state change (\emph{trigger sequence}), and one that leaks the inner state (\emph{measurement sequence}). 
Based on this notation, we introduce \ToolName, an automated tool to identify such instruction-sequence triples.
\ToolName relies on fuzzing-like techniques to combine instructions of the targeted instruction-set architecture (ISA) and analyzes whether the generated triple forms a side channel. 
\ToolName supports an efficient search scheme which can cope with side effects between different fuzzing iterations, a challenging phenomenon that is not present in most other fuzzing domains.

In contrast to CPU instruction fuzzing~\cite{Domas2017}, \ToolName does not search for undocumented instructions but instead relies on a machine-readable ISA specification. 
Such a specification exists for x86~\cite{Abel2019uops} and ARMv8~\cite{ARM2017specxml}. 
As these specifications contain all ISA extensions as well, \ToolName first reduces the candidate set to instructions that can be executed as an unprivileged user on the target CPU. 
From this candidate set, \ToolName combines instructions and tests whether they can be used as a covert channel. 
In such a case, the found triple is reported as a covert channel, and thus also as a potential side channel. 
The current proof-of-concept implementation of \ToolName is limited to finding timing-based single-instruction side channels in an unguided manner. 
However, even such a simple setup involves many challenges that require a careful design to enable finding interesting sequence triples.

We ran \ToolName for over 500 hours on 5 different Intel and AMD CPUs with microarchitectures from 2013 to 2019. 
\ToolName found both existing and novel side channels. 
The existing side channels include \FlushReload~\cite{Yarom2014Flush}, and the \styleInstruction{AVX2} side channel described by Schwarz~\etal\cite{Schwarz2019netspectre}. 
Moreover, \ToolName discovered four new side channels using the \texttt{RDRAND} and \texttt{MOVNT} instructions, as well as in the x87 floating-point and AVX vector extensions.  

In three case studies, we demonstrate that these newly identified side channels enable powerful attacks. 
Based on the findings of non-temporal moves (\texttt{MOVNT}), we show \KASLRBreak, a microarchitectural kernel-level ASLR (KASLR) break that is not mitigated by any of the hardware fixes deployed in recent microarchitectures.
We successfully evaluate \KASLRBreak on the new Intel Ice Lake and Comet Lake microarchitectures, where the performance is on par with previous microarchitectural KASLR breaks from which almost all stopped working on the newest microarchitectures.
Furthermore, with the detected side-channel leakage of \texttt{RDRAND}, we show that we can build a fast and reliable cross-core covert channel that is also applicable to the cloud. 
Our cross-core covert channel can transmit \SI{95.2}{\bit/\second} across virtual machines on the AWS cloud. 
We use these side channels as a covert channel in a Spectre and in a Meltdown attack to leak on average \SI{7.83}{\byte} in one transient window.

In addition to the practical evaluation of the side channels, we demonstrate that our new primitives can evade detection via performance counters~\cite{Chiappetta2015,Irazoqui2018mascat,Herath2015,Payer2016}, and even undermine the security of state-of-the-art proposals for secure caches~\cite{Liu2014random,Qureshi2018,Werner2019}. 
Thus, this paper shows that side channels are quite versatile, making it hard to build robust detection methods that cover all possible side channels. 
We stress that it is important to build automated tooling for analyzing the attack surface to design more effective countermeasures in the future. 
\ToolName is a first step, and even when limiting ourselves to single-instruction sequences, we show that many unknown side channels can be uncovered automatically. 

To summarize, we make the following contributions:
\begin{compactenum}
  \item We introduce an approach to automatically find timing-based microarchitectural side channels that follow a generic instruction-sequence-triple notation and develop a prototype implementation\footnote{\ToolName's source is available at \url{https://github.com/cispa/osiris}} for it.
  \item We discover 4 new side channels on Intel and AMD CPUs.
  \item We present \KASLRBreak, a microarchitectural KASLR break that works on the newest Intel microarchitectures, and a noise-resistant cross-core cross-VM covert channel that does not rely on the memory subsystem. 
  \item We analyze existing side-channel detection and prevention methods and show that they are flawed with respect to our newly discovered side channels. 
  
\end{compactenum}

\paragraph{Responsible Disclosure.}
We disclosed our findings to Intel on January 19, 2021, and they acknowledged our findings on January 22, 2021.
Moreover, we disclosed the cross-core covert channel to AMD on February 5, 2021.

\section{Background}\label{sec:background}
In this section, we provide background for this work. 

\subsection{Microarchitecture}
The microarchitecture refers to the actual implementation of an ISA. 
Typically, the microarchitecture is not fully documented, as it is transparent to the programmer. 
Hence, performance optimizations are often implemented transparently in the microarchitecture. 
As a result of the optimizations and the abstraction, there is often unintended leakage of metadata, which can be exploited in so-called microarchitectural attacks.
The most prominent microarchitectural attacks are cache-based side channels~\cite{Yarom2014Flush,Gullasch2011,Gruss2017PhD} and transient-execution attacks~\cite{Lipp2018meltdown,Kocher2019,Schwarz2019Thesis}. 

\subsection{Side- and Covert Channels}
Information is transmitted through so-called \textit{channels}.
These channels are often intended to exchange information between two entities, \eg network or inter-thread communication.
Nevertheless, some channels are unintended by the designers, \eg power consumption or response time.
Attackers can use unintended channels to transmit information between two attacker-controlled entities.
We refer to such a channel as a \textit{covert channel}.
Moreover, attackers can abuse the channel to infer inaccessible data if a victim unknowingly is the sending end.
In this case, the channel is called a \textit{side channel}.

Both side and covert channels exist in modern microarchitectures~\cite{Ge2016}.
CPU caches are probably the most popular microarchitectural components that can be abused for side or covert channels~\cite{Gruss2016Flush,Yarom2014Flush,Osvik2006,Gullasch2011}.
As CPU caches are shared among different threads and even across CPU cores, adversaries can abuse them in a wide range of attack scenarios~\cite{Gruss2015Template,Oren2015,Liu2015Last,Maurice2017Hello,Lipp2018meltdown,Kurth2020netcat}.

\subsection{Transient Execution Attacks}

As modern CPUs follow a pipeline approach, instructions might be executed out of order and are only committed to the architectural level in the correct order. %
To avoid stalling the pipeline, the processor continues precomputing even when a branch value or a jump target is unavailable, \eg due to a cache miss. 
This is enabled through several prediction mechanisms that allow speculatively executing instructions. %
When the branch target is evaluated, speculatively executed instructions are allowed to retire only in the case of correct prediction. 
Otherwise, the speculatively executed instructions are squashed. 
Instructions that are not retired but leave microarchitectural traces are called transient instructions~\cite{Lipp2018meltdown,Canella2019A,Intel2020Affected}.

Spectre~\cite{Kocher2019} is one class of transient-execution attacks exploiting speculative execution. 
By mistraining a branch predictor, an attacker can influence the transient control flow of a victim application. 
In the transient control flow, an attacker typically tries to encode application secrets into the microarchitectural state. 
Using a side channel, this encoded information is later transferred to the architectural state. 
Meltdown~\cite{Lipp2018meltdown} is another class of transient-execution attacks, exploiting the lazy handling of exceptions. 
On affected CPUs, inaccessible data is forwarded transiently before the exception is handled. %
Transient execution attacks commonly use the cache to encode leaked secrets~\cite{Lipp2018meltdown, Kocher2019, Koruyeh2018spectre5, Maisuradze2018spectre5, Canella2019A} but can also use other side channels~\cite{Schwarz2019netspectre, Bhattacharyya2019, Schwarz2019STL, Lipp2020takeaway}.

\subsection{Fuzzing}
Fuzzing is a software testing technique that aims at finding bugs in software applications~\cite{Aschermann2019redqueen, Chen2018angora, Peng2019tfuzz, Rawat2017, Stephens2016}. 
A fuzzer typically generates a large number of test inputs and monitors software execution over these inputs to detect faulty behavior. 
Due to the huge input space, fuzzers typically search for inputs with a high probability of triggering a bug while avoiding uninteresting input.
Fuzzers usually follow one of two different approaches for generating input~\cite{Aschermann2019redqueen, Blazytko2019grimoire}.
Mutation-based fuzzers usually start with an initial set of inputs (seeds), then generate further test input by applying mutations, \eg splicing or bit flipping~\cite{Eddington_peach, Aschermann2019redqueen, Hocevar2015zzuf}.
Grammar-based fuzzers exploit existing input specifications to generate a model of the expected input format. 
Based on this model, the fuzzer efficiently generates accepted input~\cite{Blazytko2019grimoire, Han2019codealchemist, Padhye2018zest}. 
Moreover, fuzzing approaches can be clustered in two classes based on how they generate new or mutated input. 
While blind fuzzing randomly generates input based on a grammar of predefined mutations~\cite{Helin_radamsa, Eddington_peach}, guided fuzzing uses the current execution to guide the generation of new input. 
These techniques aim to maximize a given metric~\cite{Zalewski2014afl, Aschermann2019redqueen, Peng2019tfuzz, Chen2018angora}. 

Most research efforts on fuzzing target software applications. 
Nonetheless, hardware fuzzing is becoming increasingly popular~\cite{Gras2020,Moghimi2020medusa,Domas2017}.
Sandsifter~\cite{Domas2017} presents a search algorithm that allows efficiently finding undocumented x86 instructions. 
It applies byte-code mutation to generate new instructions and checks whether the processor can decode the generated instructions.
ABSynthe~\cite{Gras2020} allows automatically synthesizing a contention-based side channel for a target program.
It uses fuzzing to find instruction sequences that generate distinguishable contention on secret-dependent code execution.
Mutation parameters in ABSynthe include instruction building blocks, repetition number, and use of memory barrier.
Hardware fuzzing has also been utilized to improve existing Meltdown attacks~\cite{Xiao2019Speechminer} or find new variants of these attacks~\cite{Moghimi2020medusa}, automate the search for Spectre gadgets~\cite{Tol2020FastSpec}, and identify cross-core transient-execution attacks~\cite{Ragab2021}.

\section{High-level Overview of \ToolName}\label{sec:design}
In this section, we introduce a notation that captures timing-based side channels based on \emph{instruction-sequence triples} (\Cref{sec:design:notation}) before we describe the design of \ToolName.
Side channels not exploitable via timing differences are out of scope for \ToolName. 
We discuss challenges when using this new notation to find side channels (\Cref{sec:design:challenges}).
Finally, we showcase the big picture of our fuzzing framework (\Cref{sec:design:fuzzer}).

\subsection{Side-Channel Notation}
\label{sec:design:notation}

\begin{table}[t]
\begin{center}
\caption{Existing timing-based side channels mapped to sequence triples and whether our approach can find it (\protect\circletfill) or cannot find it (\protect\circlet). Reasons for failure are that multiple instructions are required (\protect\reasonMI), side channel only works across hardware threads (\protect\reasonHT), or specific operands are required (\protect\reasonOP).
}
\label{tab:three-sequence}
\adjustbox{max width=\hsize}{
 \begin{tabular}{llllll}
\toprule
 \textbf{Side channel} & \textbf{\ReSeq} & \textbf{\TrSeq} & \textbf{\MeSeq}  & \ToolName & Reason \\
 \midrule
 \styleInstruction{AVX}~\cite{Schwarz2019netspectre} & sleep & \styleInstruction{AVX2} instr. & \styleInstruction{AVX2} instr.  & \circletfill & \\
  \FlushReload~\cite{Yarom2014Flush} & \styleInstruction{CLFLUSH} & mem.~access & mem.~access  & \circletfill & \\
  \FlushFlush~\cite{Gruss2016Flush} & \styleInstruction{CLFLUSH} & mem.~access & \styleInstruction{CLFLUSH}  & \circletfill & \\
  Flush+Prefetch~\cite{Gruss2016Prefetch} & \styleInstruction{CLFLUSH} & {mem.~access} & \styleInstruction{PREFETCH}  & \circletfill & \\
  BranchScope~\cite{Evtyushkin2018} & cond.~jump & cond.~jump & cond.~jump  & \circletfill & \\
  \EvictReload~\cite{Percival2005} & {mem.~accesses} & {mem.~access} & {mem.~access} & \circlet & \reasonMI, (\reasonOP) \\
  \EvictTime~\cite{Osvik2006} & {mem.~accesses} & {mem.~access} & {mem.~access} & \circlet & \reasonMI, (\reasonOP) \\
  \PrimeProbe~\cite{Percival2005} & {mem.~accesses} & {mem.~access} & {mem.~accesses}  & \circlet & \reasonMI, \reasonOP \\
  Reload+Refresh~\cite{Briongos2020reload} & {mem.~accesses} & {mem.~access} & {mem.~accesses}  & \circlet & \reasonMI, \reasonOP \\
  Collide+Probe~\cite{Lipp2020takeaway} & {mem.~access} & {mem.~access} & {mem.~access}  & \circlet & \reasonOP \\
  DRAMA~\cite{Pessl2016} & mem. access & mem. access & mem. access & \circlet & \reasonOP \\
  Port contention~\cite{Aldaya2018} & sleep & execute & execute (same HT) & \circlet & \reasonHT \\
  \bottomrule
 \end{tabular}
 }
 \end{center}
\end{table}

For detecting side channels, we first focus on detecting covert channels, as every side channel can also be used as a covert channel. 
Regardless whether timing-based covert channels are used as side channels or as covert channels in transient-execution attacks, they follow these three steps: 

(1) In the first step, the attacker brings a microarchitectural component, abused by the attack, into a known state. 
For example, the attacker might flush or evict a cache line (\eg \FlushReload, \PrimeProbe, \EvictReload) or power down the \styleInstruction{AVX2} unit. 
We call this known state the \emph{reset state} (\sZero). 
We call a sequence of instructions that causes a transition to \sZero a \emph{reset sequence} (\ReSeq).

(2) In the second step, the victim (or the sending end) changes the state of the abused microarchitectural component based on a secret.
The victim might cache a value depending on the secret, or power up the \styleInstruction{AVX2} unit by executing an \styleInstruction{AVX2} instruction. 
We call the new state the \emph{trigger state} (\sOne). 
We call a sequence of instructions causing a transition to \sOne a \emph{trigger sequence} (\TrSeq).

(3) Finally, the attacker tries to extract the secret value by checking whether the abused component is in the reset state \sZero or the trigger state \sOne. 
This is typically done by measuring the execution time of a particular instruction sequence, which we call the \emph{measurement sequence} (\MeSeq).
The measurement sequence may---in fact, typically \emph{does}---have side effects beyond measuring, \ie it also influences the state.

\Cref{tab:three-sequence} shows examples of these three instruction sequences for several known side channels.
For example, \FlushReload uses \texttt{CLFLUSH} as the reset, and memory accesses (\eg via \texttt{MOV}) as trigger and measurement sequences.
The careful reader will notice that existing side channels often do not require instruction \emph{sequences}, but just a single instruction per step---a simplification that we will leverage ourselves later.

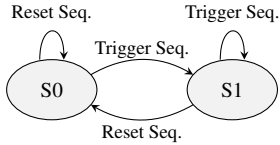
\begin{figure}
  \centering
  \input{images/states.tikz}
  \caption{State machine representing different microarchitectural states and transitions between them.}
	\label{fig:SM}
\end{figure}
\cref{fig:SM} shows a state machine representing the relation between the three steps of an attack and the different microarchitectural states of the abused component.
These two states could represent an abstraction over possibly more complex states of the component, \eg different cache levels. 
However, to mount a side-channel attack, it is sufficient to distinguish and transit between two states only.

\subsection{Challenges of Side-Channel Fuzzing}
\label{sec:design:challenges}

Based on this notation, we design \ToolName, a fuzzer that aims to automatically find new microarchitectural side channels.
The overall idea is to generate inputs, \ie instruction-sequence triples, and then detect whether such a triple forms a side channel. 
For this, \ToolName executes a triple and measures the execution time of the measurement sequence.
At an abstract level, we compare timings \emph{with} and \emph{without} prior execution of the trigger sequence.
Large timing differences hint at side channels.
While the overall idea is intuitive, several challenges complicate the search:

\paragraph{Unknown Sequences.}
First, as we aim for \emph{novel} side channels, we cannot assume \emph{a priori} knowledge of valid reset, trigger, or measurement sequences.
This poses a significant challenge to fuzzing, as we have to fuzz all three inputs without knowing their relations.
We are unaware whether an instruction sequence actually is a reset, trigger, or measure sequence.
Even if we find a sequence (\eg a trigger), we do not know which counterparts are required for the other two sequences (\eg corresponding reset and measurement sequences).

\paragraph{Unknown Side Effects.}
Second, sequences on their own may have undesired side effects, such as measurement sequences that change the state.
For example, memory accesses within the measurement sequence do not only passively observe the memory access time, but they also change the cache state.
This implies that our state diagram becomes more complex, as measurement sequences may in fact act as triggers themselves.
If we had a valid reset sequence, this would not be a problem, as we could revert this change.
However, as mentioned above, we do not know the corresponding reset sequence, and therefore have to mitigate this problem conceptually.

\paragraph{Dirty State.}
Third, in the interest of efficiency, we want to fuzz as fast as possible.
This, unfortunately, means that a subsequent sequence triple may inherit a dirty, non-pristine state from its successor.
For example, if the first triple contains a memory access, the triple executed after that likely inherits the cache state.
In other words, we cannot assume that sequence triples run in isolation.
They do affect each other.

\paragraph{Generality.}
Fourth, we want to be as generic as possible and cover the entire instruction set of a given ISA.
That is, instead of testing just a few popular instructions, we would like to explore the entire range of instructions and their combinations.
To this end, we not only require knowledge of all instructions but also a semantic understanding of an instruction's syntax, such as its operands and their types.

\paragraph{Indistinguishability.}
Finally, executing similar instructions inevitably leads to similar, if not indistinguishable\footnote{Indistinguishable side channels are those which lead to the same attacker observation on system states.}, side-channel candidates.
In fact, we create thousands of sequence triples, many of which are close to each other.
For example, with reference to known side channels, dozens of instructions use vector operations to power up the AVX unit.
However, regardless of which instruction is executed, more or less the same side channel is found.
\cref{sec:implementation} elaborates on how we solved these challenges for \ToolName. 

\subsection{Big Picture}
\label{sec:design:fuzzer}

\begin{figure*}[t]
  \centering
    \input{images/overview.tikz}
  \caption{Overview of \ToolName. 
  The offline phase extracts available instructions from a machine-readable ISA description. 
  The first phase generates sequence triples from these instructions. 
  The execution phase measures their execution times and forwards triples with timing differences to the confirmation phase. 
  If the timing difference persists on randomized execution of the triple, it is considered a side channel and forwarded to the clustering phase, which categorizes the triple and creates the final report.}
	\label{fig:implementation}
\end{figure*}
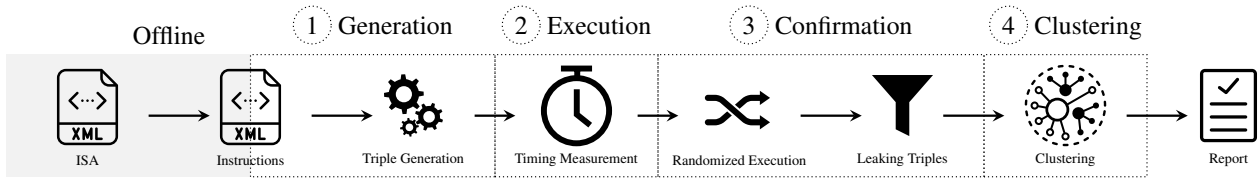

\Cref{fig:implementation} shows the big picture of \ToolName, a fuzzer that tackles these challenges.
In step~\circled{1}, the \emph{code generation stage}, we fuzz potential instruction sequences, \ie  triples of \ReSeq, \TrSeq, and \MeSeq.
These sequences are generated from a machine-readable specification of the targeted architectures.
The generated triples are then forwarded to step~\circled{2}, the \emph{code execution stage}.
Here, the generated triples are executed in a special order (at least) twice---once including the trigger (\emph{hot path}), and once without (\emph{cold path}).
We time the measurement sequence (\MeSeq) of both paths to see if the trigger sequence (\TrSeq) causes timing differences.
The timing difference is then processed in step~\circled{3}.
This \emph{result confirmation stage} interprets a large timing difference as the first indicator of whether a given triple constitutes a side channel candidate.
On top of this, to address many of the problems as mentioned earlier, there are additional validation routines that sort out actual side channels from wrong candidates.
For example, we check whether (i) the reset sequence has any effect at all to exclude a bad triple combination, and (ii) a different fuzzing order confirms the result.
Finally, in step~\circled{4}, we feed the list of confirmed side channels to the \emph{clustering stage}.
This step clusters similar, indistinguishable side channels, to ease further analyses of the side channels.

\section{Design and Implementation}\label{sec:implementation}

Next, we discuss the implementation of \ToolName for the x86 ISA and how we solved the challenges enumerated in \Cref{sec:design:challenges}. 
While we chose to implement and evaluate our fuzzer on this architecture, the overall design is equally applicable to processors that use a different instruction set, e.g., ARM processors. 
In the following, we present the implementation details for the four stages outlined in \Cref{fig:implementation}.

\subsection{Code Generation Stage}
The goal of the code generation stage is to produce triples of assembly instruction sequences (a reset sequence \ReSeq, a trigger sequence \TrSeq, and a measurement sequence \MeSeq). 
Since we are not aware of a clear feedback mechanism that can guide the creation of sequence triples, we opted for the creation of random x86 instructions.
To bootstrap the code generation, we employ a grammar based on a machine-readable specification of x86 instructions. 
The code generation involves two phases: (1)~an offline phase where all supported instruction sequences are generated, and (2)~an online phase performing the creation of triples. 
The offline phase is executed once for each ISA and consists of instruction creation and machine-code file generation. 
The online phase is executed repeatedly for each run of the fuzzing process.

\subsubsection{Offline Phase}
The output of the offline phase is an assembly file containing all possible instruction variants for the target ISA. 
This file is generated once and reduces the overhead required for generating and assembling instructions during runtime. 

\paragraph{Generation of Raw Instructions.}
The first task is the generation of all valid x86 instructions. 
To achieve this, we leverage a machine-readable x86 instruction variant list from uops.info~\cite{Abel2019uops}. 
This list extends Intel's XED iForm\footnote{\url{https://intelxed.github.io/ref-manual/xed-iform-enum_8h.html}} with additional attributes, \eg effective operand size, resulting in a large number of instruction variants per instruction. 
For example, this list provides 35 variants for the mnemonic \styleInstruction{MOV} and 26 variants for the mnemonic \styleInstruction{XOR}, summing up to \SIx{14039} x86 instruction variants overall.
The list also contains comprehensive information about each instruction variant, \eg extension or category, that we later use for the clustering.

\paragraph{Creation of the Machine Code.}
The second task is assembling the instructions to machine code.
We try to reduce the number of instructions by treating all registers as equivalent, \ie \ToolName does not generate the instruction with all possible register combinations.
\ToolName, w.l.o.g, relies on a fixed set of registers as operands for each instruction.
We also exclude instructions that change the control flow (\eg \styleInstruction{RET}, \styleInstruction{JMP}) as they may lead to an irrecoverable state.
As branches have been studied extensively for microarchitectural attacks~\cite{Acicmez2007new,Aciicmez2007e,Aciicmez2007predicting,Evtyushkin2015,Evtyushkin2016ASLR,Lee2017Inferring,Evtyushkin2018,Kocher2019}, we do not assume that \ToolName would find any new side channels for these instructions. 
Finally, we add a pseudo-instruction that allows idling the CPU for a certain period of time. 
This instruction is required to reset components that are based on power-saving features of the CPU, \eg the \styleInstruction{AVX2} SIMD unit.
For each assembled instruction, the file also stores a set of attributes, \eg the ISA extension or instruction category, that are used in the clustering phase.

\begin{table}
\begin{center}
  \caption{Faulting instructions on Intel \styleArchitecture{Core i7-9750H}.}
  \label{tbl:faults}
  \adjustbox{max width=\hsize}{
  \begin{tabular}{lr}
   \toprule
    \textbf{Signal}    & \textbf{Number of Occurrences} \\
    \midrule
    Segmentation fault (\textit{SIGSEGV})          &     118       \\    
    Floating-point exception (\textit{SIGFPE})           &     22        \\    
    Illegal instruction (\textit{SIGILL})           &     \SIx{10508}     \\    
    Debug instruction (\textit{SIGTRAP})          &     1         \\   
    \bottomrule
  \end{tabular}
  }
\end{center}
\end{table}

\subsubsection{Online Phase}
When starting \ToolName on a machine, the online phase first removes instructions that are not supported on the microarchitecture, and then generates all possible sequence triples.

\paragraph{Cleanup of Machine-Code File.}
The first task is the cleanup of the machine-code file generated in the offline phase.
This is required since the generated machine-code file contains instruction variants for the entire x86 ISA, including all extensions. 
Hence, it contains a significant number of illegal instructions for a given microarchitecture. %
Moreover, the file may also include instructions that generate faults when executed by our framework, \eg privileged instructions.
The cleanup process is done by executing all instructions once and maintaining a list of all the instructions that terminated normally. 
This process reduces the number of instructions in the machine-code file considerably. 
For example, the number of user-executable instructions for an \styleArchitecture{Intel Core i7-9750H} is \SIx{3390}, \ie \SI{24.1}{\percent} of the instruction variants initially generated in the offline phase. 
\cref{tbl:faults} shows the distribution of faults generated in the cleanup process for this processor.
The majority of the faults (\SI{98.7}{\percent}) are illegal-instruction faults, \ie the instruction is not supported at all or not in user space. 

\paragraph{Generation of Sequence Triples.}
The second task is the generation of sequence triples from the list of executable instructions that are forwarded to the code execution stage.
We exploit three observations that allow reducing the complexity of this task as well as the overhead of the fuzzing process:
\begin{compactenum}
    \item Most existing non-eviction-based side channels require only one instruction in each of the sequences.
    \item Idling the processor is used only as a reset sequence.
    \item Trigger and measurement sequences may be formed of exactly the same instruction.
\end{compactenum}
Consequently, in our implementation, the triples are generated by considering all possible combinations of single instructions, where the sleep pseudo-instruction is only used as a reset sequence. 
While our framework is easily extensible to support multi-instruction sequences, the search space quickly explodes---a topic we thus leave open to future work.

\subsection{Code Execution Stage}
The goal of the code execution stage is to execute generated input triples and analyze their outcome, \ie determine whether an executed triple forms a side channel.

\paragraph{Environment.} 
The triple is executed within the process of \ToolName to not suffer from the additional overhead of process creation. 
To reduce external influences, such as interrupts, \ToolName relies on the operating system to reduce any noise.
First, the operating system ensures that there are no core transitions that influence the measurement by pinning the execution of the triple to a dedicated CPU core. 
Additionally, this entire physical core is isolated to ensure that the code is unlikely to be interrupted, \eg by the scheduler or hardware interrupts. 

\paragraph{Setup.}
To measure the execution time of a triple, it is placed on a dedicated page in the address space between a special prolog and epilog.
The prolog is responsible for saving all callee-saved registers according to the x86-64 System~V ABI~2. 
The prolog furthermore ensures that the triple has one page of scratch space on the stack. 
Thus, there is no corruption if any of the instructions in the triple modifies the stack, \eg the \styleInstruction{POP} instruction. 
Furthermore, the prolog initializes all registers that are used as memory operands to the address of a zero-initialized writable data page. 
This prevents corrupting the memory of \ToolName and ensures that executed instructions access the same memory page.
Note that the zero-filled page is always the same, and the framework resets this page for every tested triple.
The epilog is responsible for restoring the registers and the stack state, ensuring that any architectural change is reverted.
Moreover, signal handlers are registered for all possible signals that can arise from executing an instruction, \eg \styleSignal{SIGSEGV}. 
These handlers abort the execution of the current triple and restore a clean state for \ToolName.
Finally, we abstain from parallelization, as this could lead to unexpected interferences in shared CPU resources. 

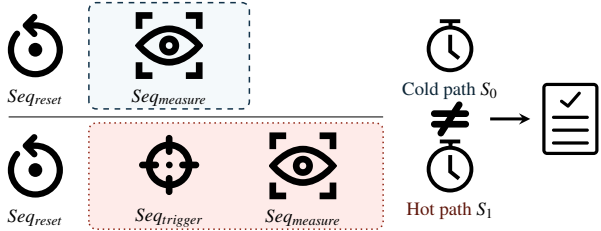
\begin{figure}
 \input{images/measure.tikz}
 \caption{The execution stage receives the triple and executes \MeSeq (cold path) and \TrSeq, \MeSeq (hot path) after \ReSeq.
 Timing differences for the two paths are reported. }
 \label{fig:execution-stage}
\end{figure}

\paragraph{Measurement.}
Once the triple is prepared, \ToolName executes the generated sequence twice, once with the trigger sequence \TrSeq (hot path) and once without (cold path), as illustrated in \Cref{fig:execution-stage}. 
In both cases, the execution time of the measurement sequence \MeSeq is measured. 
This code aims to detect the existence of a side channel by observing timing differences in the measurement instruction, depending on whether or not a trigger was used.
A significant difference between the two measurements indicates a candidate side channel that is then forwarded to the confirmation stage. 
To ensure precise time measurement and no unintentional dependency on the timing measurement itself, we add serializing and memory-ordering instructions around the measured code.

\subsection{Result Confirmation Stage}

The goal of the confirmation stage is to validate if a triple reported by the execution stage is an exploitable side channel. 
To confirm or refute these candidates, \ToolName further analyzes the identified triples to rule out other side effects that could have led to the detected timing difference. 
Such side effects include unreliable reset sequences or a dirty state caused by previous execution (\cf \Cref{sec:design:challenges}). 
To eliminate non-promising candidates, we foresee the following mechanisms.

\paragraph{Repeated Execution.}
External factors, such as power-state changes or interrupts, can induce timing differences.
To rule out such cases, \ToolName executes the hot path and the cold path (\cf \Cref{sec:design:fuzzer}) over a predefined number of runs to compare the median of the timings for the two cases.
In particular, this check is passed if the difference between the two medians is greater than a predefined threshold.
The number of measurements is a parameter that allows setting a tradeoff between precision and runtime. 
While a high number of repetitions takes longer, it increases the confidence in the result, as external influences are statistically independent and thus average out. 
Too few repetitions reduce the confidence in the accuracy of the reported results, leading to false positives. 

\paragraph{Non-Functional Reset Sequences.}
The initially observed timing difference may result from different sequence combinations leading to the desired state without actually performing the required transition. 
For example, consider a faulty reset sequence \ReSeq that does not reset the state to \sZero. 
A timing difference would still be detected by the first check if the test started in a state \sZero. 
To ensure the correct functionality of \ReSeq, \ToolName measures the execution time of \MeSeq after the execution of \ReSeq. 
It then measures the timing after the execution of \TrSeq followed by \ReSeq. 
A negligible difference between the two measurements indicates that \ReSeq actually resets the state to \sZero when triggered to \sOne by \TrSeq. 
The check also implies that the state change observed in the first check must be caused by executing \TrSeq. 
Consequently, the input formed of the sequence triple allows reaching the target, \ie it represents a potential side-channel.

\paragraph{Triple Reordering.}
\ToolName executes all generated triples shortly after another. We may therefore experience undesired edge cases caused by dirty microarchitectural states and side effects caused by prior executions. We therefore test each sequence multiple times (twice in our evaluation), each time randomizing the order in which we test the fuzzed triples. We then ignore triples that do show discontinuous behavior in all tested permutations. This reordering ensures that we have a negligible probability that two given sequence triples are executed directly after each other in both runs, hence lowering the chances of repetitive dirty states being carried over. 

\paragraph{Applicability in Transient Execution.}
\ToolName also allows detecting whether a side channel can be used as covert channels for transient-execution attacks.
To test the transient behavior of the side channel, \ToolName executes \TrSeq speculatively using Retpoline as shown in previous work~\cite{Wong2018specpoline,Stecklina2018}. 
We opted for this variant as it has a perfect misspeculation rate requiring no mistraining of any branch predictors~\cite{Wong2018specpoline}.
\ToolName allows to optionally enable this behavior in the confirmation stage.

\subsection{Clustering Stage}
Different sequence triples can lead to the detection of the same side channel. 
For example, for cache-based side channels, every instruction that accesses a memory address can act both as trigger and as measurement sequence. 
Due to the CISC nature of x86, many instructions explicitly (\eg \texttt{ADD}) or implicitly (\eg \texttt{PUSH}) access memory. 
Additionally, every instruction that flushes this address acts as a reset sequence.
Similarly, in the \styleInstruction{AVX2} side channel, different \styleInstruction{AVX2} instructions can act both as trigger and as measurement sequence.

In the clustering stage, \ToolName aims at clustering the input forwarded from the code execution stage into groups that represent different side channels. 
To achieve this, we can base our clustering on various properties of the involved instruction sequences. 
Examples of instruction properties include the instruction's extension, memory behavior, and the general instruction category (\eg arithmetic or logical).
Additionally, our tests showed that the timing difference tends to be an important clustering property.
This procedure assumes that similar side channels show similarities in the properties of the corresponding instructions. 
We identify two categories of properties that can be used for clustering, as outlined next.

\paragraph{Static Properties.}
Triples can be classified based on properties of the contained instructions, such as the instruction category (\eg \emph{arithmetic} or \emph{logical}) or the instruction extension (\eg \styleInstruction{AVX2} or \styleInstruction{x87-FPU}).
As this information is propagated from the instructions to the clustering phase, \ToolName fundamentally relies on this information for clustering. 
The clustering stage clusters the reported triples based on the instruction set extension of \TrSeq and \MeSeq. 
The intuition behind this clustering is that instruction-set extensions are strong indicators for the underlying microarchitectural root cause. %
Although this process cannot remove all duplicates, it significantly reduces the number of reported triples, thus, facilitating further analysis of the side channels.

\paragraph{Dynamic Properties.}
In addition to the static properties of instructions, it is also possible to cluster triples based on their dynamic effects. 
One of the dynamic properties \ToolName supports for clustering is the observed timing difference. 
If multiple triples lead to the same timing difference, the root cause is likely the same, \ie access-time differences when accessing cached and uncached memory. 
Additionally, the clustering stage may cluster the triples based on their cache behavior. 
As shown by Moghimi~\etal\cite{Moghimi2020medusa}, performance counters can be used for clustering triples. 
By executing triples while recording performance counters, it is possible to dynamically observe which parts of the microarchitecture are active. 
This can also help to identify the root cause easier.

\section{Results}\label{sec:results}
In this section, we evaluate the design choices of \ToolName based on the prototype implementation described in \cref{sec:implementation}.

\subsection{Evaluation Setup}
We perform the fuzzing on 5 different CPUs and evaluate the case studies based on our results on a more extensive set of CPUs (\cf \Cref{tab:kaslr-cpus} and \Cref{tab:rdrand-cpus}).
We use a laptop with an \styleArchitecture{Intel Core i7-9750H} (Coffee Lake), and 4 desktop machines with an \styleArchitecture{Intel Core i7-9700K} (Coffee Lake), \styleArchitecture{Intel Core i5-4690} (Haswell), \styleArchitecture{AMD Ryzen 5 2500U} (Zen), and \styleArchitecture{AMD Ryzen 5 3550H} (Zen+).
All systems run Ubuntu or Arch Linux.

\subsection{Performance}
Before demonstrating \ToolName's ability to find side channels, we evaluate its performance, \ie the number of triples tested per second. 
To measure this throughput, we first use the same instruction sequence for \TrSeq and \MeSeq.
For the first measurement, we exclude the pseudo sleep instruction, as it---by construction---biases the code execution time.
We only report the throughput for the oldest processor, \ie the \styleArchitecture{Intel Core i5-4690}.
For this microarchitecture, there are \SIx{3377} instructions (after cleanup), leading to a total of \SIx{3377}$^2$ = \SIx{11404129} sequence triples.
A full fuzzing run terminated in just \SI{41}{\second}, resulting in a throughput of \SIx{278149} triples per second.
To identify the bottleneck of our framework, we increased the number of repetitions of each triple from 1 to 10, \ie executed more code.
In this experiment, the fuzzer took \SI{127}{\second} to complete (\SIx{89796} triples per second), resulting in a runtime increase by factor 3 only.

When including the pseudo sleep instruction, the overall runtime grows to \SI{56}{\second} and \SI{271}{\second} for 1 and 10 repetitions, respectively.
That is, the throughput reduces to \SIx{202370} triples per second (or \SIx{42044} for 10 repetitions).
This is a \SI{37}{\percent} slowdown compared to the first run that excluded sleeping.
Intuitively, sleeps imply that the fuzzer spends more time executing code.
This explains the stronger impact of the actual code execution on the overall throughput compared to code generation.
Increasing the number of repetitions by 10x, therefore, decreases the number of tested triples by a factor of 4.8.
For the actual fuzzing run, \TrSeq and \MeSeq are different. 
Hence, the number of sequence triples increases to \SIx{3377}$^3$ = \SIx{38511743633}, leading to a runtime of nearly 5 days. 

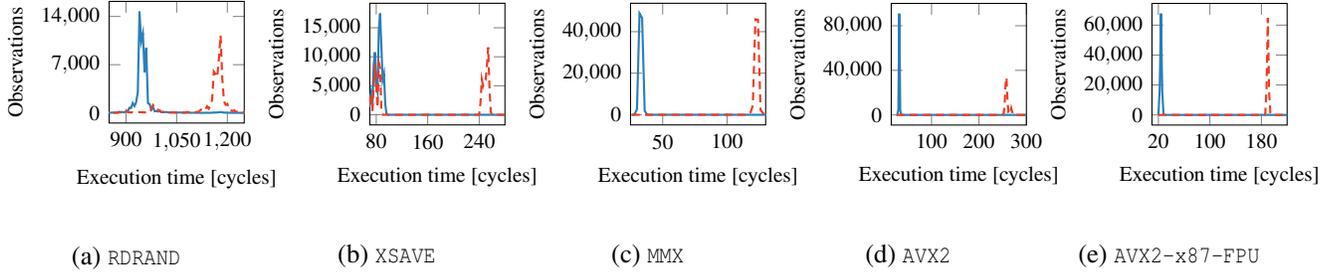
\begin{figure*}[htp]
\centering
\begin{subfigure}{.19\hsize}
\input{images/rand.tikz}
\caption{\styleInstruction{RDRAND}}
\label{fig:rdrand}
\end{subfigure}
\begin{subfigure}{.19\hsize}
\input{images/xsave.tikz}
\caption{\styleInstruction{XSAVE}}
\label{fig:xsave}
\end{subfigure}
\begin{subfigure}{.19\hsize}
\input{images/mmx.tikz}
\caption{\styleInstruction{MMX}}
\label{fig:mmx}
\end{subfigure}
\begin{subfigure}{.19\hsize}
\input{images/avx-pause.tikz}
\caption{\styleInstruction{AVX2}}
\label{fig:avx2}
\end{subfigure}
\begin{subfigure}{.19\hsize}
\input{images/avx-x87.tikz}
\caption{\styleInstruction{AVX2-x87-FPU}}
\label{fig:avx2-x87}
\end{subfigure}
\caption{Histograms of \MeSeq execution time depending on whether \TrSeq was executed (solid blue) or not (dashed red).}
\label{fig:histograms}
\end{figure*}

\subsection{Clustering}
On the tested microarchitectures, \ToolName successfully clustered the reported instances into fewer than 30 clusters.
On the \styleArchitecture{Intel i7-9750H}, the \SI{68597} reported side channels were first clustered into 186 clusters. 
To further reduce the number of clusters caused by one side-channel variant, \ToolName also provides the clustering based only on \TrSeq and \MeSeq, as these sequences contain the instructions causing the leakage. 
Based on these two sequences, the number of clusters is only 16. 
\Cref{tbl:clusters} (\Cref{sec:appendix:cluster}) shows the numbers for other CPUs. 

\subsection{Rediscovering Known Side Channels}
A typical test for software fuzzer is the rediscovery of old bugs, \eg by searching for vulnerabilities in poorly tested software, checking for well-known CVEs, or uncovering bugs reported by prior work.
\ToolName also rediscovered two well-known side channels, \FlushReload~\cite{Yarom2014Flush} and the \styleInstruction{AVX2}-based side channel~\cite{Schwarz2019netspectre}, as described in the following. 
\Cref{sec:discussion} discusses some of the known side channels \ToolName did not rediscover and provides the reason for that. 

\paragraph{\FlushReload-Based Side Channel.}
\ToolName detects a total of \SIx{18799} triples that can be classified as a variant of \FlushReload. 
These triples have in common that \ReSeq is in either \styleInstruction{CLFLUSH} or \styleInstruction{CLFLUSHOPT}, and \TrSeq is some kind of memory load. 
Interestingly, we also found a new variant of \FlushReload that uses \styleInstruction{MOVNTDQ} as \ReSeq. 
This store instruction with a non-temporal hint also evicts the accessed memory address from the cache~\cite{Intel_opt}.

Arguably, in a practical attack, this is not very useful, as writable shared memory is typically not a target for \FlushReload. 
However, in the case of transient-execution attacks, where an attacker often uses \FlushReload as a covert channel to transfer the leaked data from the microarchitectural domain to the architectural domain, this alternative flushing method is indeed useful. 
In \Cref{sec:study:transient}, we show that the MOVNT-based \FlushReload can increase the leakage from 3 to 7.83 bytes per transient window for Meltdown-type attacks, reducing the impact of the \FlushReload part that is often the bottleneck.

\paragraph{\styleInstruction{AVX2}-Based Side Channel.}
\ToolName also found 514 instances of the AVX-based side channel~\cite{Schwarz2019netspectre}. 
For this side channel, the \TrSeq and \MeSeq contain \styleInstruction{AVX2} or \styleInstruction{AVX512} instructions, and \ReSeq is simply idling. 
According to Schwarz~\etal\cite{Schwarz2019netspectre}, a busy-wait executing for around \SIx{2700000} cycles would power down the \styleInstruction{AVX2} SIMD unit. 
However, our manual tests showed that a busy wait of \SIx{8000} cycles is, in fact, sufficient.

Interestingly, we also observed during the manual inspection a variant of the \styleInstruction{AVX2} side channel that contains the \styleInstruction{PAUSE} in its \ReSeq. 
\cref{fig:avx2} visualizes the behavior of this new variant for \SIx{200000} executions.
As shown in the figure, this variant is, in fact, more stable than the variant based on busy wait.
In particular, we observed a difference of \SIx{226} cycles between the medians of the two distributions, which is twice the difference for triples that have a busy-wait as \ReSeq.

\begin{table*}
\centering
  \caption{Overview of the novel side channels.}
  \label{tbl:new}
  \resizebox{\textwidth}{!}{
  \begin{tabular}{llllr}
  \toprule
\textbf{Side Channel Name} & \textbf{Example \TrSeq{}} & \textbf{Example \MeSeq{}} & \textbf{Example \ReSeq{}} & \textbf{Timing Diff.} \\
\midrule
RDRAND  & \styleInstruction{RDRAND}  & \styleInstruction{RDRAND}    & Sleep Pseudo-Inst.  & 228 cycles \\
XSAVE   & \styleInstruction{XSAVE [R8]}  & \styleInstruction{XSAVE [R8]}    & \styleInstruction{LAR ECX, EDX}  & 158 cycles \\
MMX-x87-FPU & \styleInstruction{PHADDD MM1, [R8]}  & \styleInstruction{PHADDD MM1, [R8]}    & \styleInstruction{FLDLN2}  & 90 cycles \\
AVX2-x87-FPU & \styleInstruction{VDMADD132PD YMM1, YMM2, [R8]}  & \styleInstruction{VFMADD132PD YMM1, YMM2, [R8]}    & \styleInstruction{FISTP [R8]}  & 166 cycles \\
\bottomrule
  \end{tabular}
  }
\end{table*}
\subsection{Finding Novel Side Channels}
To demonstrate the effectiveness of our fuzzer, we tested its ability to uncover new side channels. 
After running our fuzzer for 21 days, we automatically uncovered 4 different, previously unknown side channels.  
\cref{tbl:new} shows an overview of the reported side channels. 
In the following, we briefly present each of these side channels.

\paragraph{\styleInstruction{RDRAND}-Based Side Channel.}
This side channel consists of triples having the \styleInstruction{RDRAND} instructions in both \TrSeq and \MeSeq, and the sleep pseudo-instruction in \ReSeq.
\cref{fig:rdrand} visualizes the behavior of this side channel for \SIx{200000} executions. 
We observed a difference of 228 cycles between the medians of the two distributions.
Setting a simple threshold to the average of these two medians leads to a success rate of \SI{84.28}{\percent} when attempting to distinguish between the two states \sZero and \sOne.
While it is unlikely that detecting the execution of the \texttt{RDRAND} instruction leads to a side-channel attack, we demonstrate in \Cref{sec:study:covert} that this finding can be used for a stealthy cross-core covert channel. 

\paragraph{\styleInstruction{XSAVE}-Based Side Channel.}
This side channel consists of triples having the \styleInstruction{XSAVE} or \styleInstruction{XSAVE64} instructions in both \TrSeq and \MeSeq.
For this side channel, \ReSeq can contain various instructions. 
However, we distinguish between two variants: 
(1)~a non-transient variant that contains \styleInstruction{LSL}, \styleInstruction{RDRAND}, \styleInstruction{LAR}, \styleInstruction{FLD}, \styleInstruction{FXRSTOR64}, or \styleInstruction{FXSAVE64} instructions in \ReSeq; and
(2)~a transient variant that contains \styleInstruction{XSAVEOPT} instruction in addition to most \styleInstruction{x87-FPU} instructions.
 
\cref{fig:xsave} visualizes the behaviour for \SI{200000} executions of a triple formed of \styleInstruction{XSAVE [R8]} in both \TrSeq and \MeSeq, and \styleInstruction{LAR ECX, EDX} in \ReSeq.
We observed a difference of 158 cycles between the medians of the two distributions.
Using the average of the two medians as threshold leads to a rather unstable behaviour, though.
We observe a success rate of only \SI{75.10}{\percent} when attempting to distinguish between the two states \sZero and \sOne.

\paragraph{\styleInstruction{MMX} Combined with \styleInstruction{x87-FPU}.}
This side channel consists of triples having the \styleInstruction{MMX} instructions in both \TrSeq and \MeSeq, and \styleInstruction{x87-FPU} in \ReSeq.
\cref{fig:mmx} shows the histogram for \SI{200000} executions of the triples. 
The reported triples have a time measurement difference of 90 cycles in the median.
We could reliably distinguish between the states \sZero and \sOne with an accuracy of \SI{99.99}{\percent}.

\paragraph{\styleInstruction{AVX2} Combined with \styleInstruction{x87-FPU}.}
This side channel consists of triples having the \styleInstruction{AVX}, \styleInstruction{AVX2}, \styleInstruction{AVX512}, \styleInstruction{FMA}, or \styleInstruction{F16C} instructions in both \TrSeq and \MeSeq, and \styleInstruction{x87-FPU} in \ReSeq.
The reported triples have a time measurement difference in the interval of 72 to 208 cycles.

\Cref{fig:avx2} visualizes the behavior for \SIx{200000} executions of a triple formed of \styleInstruction{VFMADD132PD YMM1, YMM2, [R8]} in both \TrSeq and \MeSeq, and \styleInstruction{FISTP [R8]} in \ReSeq.
We observe a difference of 166 cycles between the medians of the two distributions.
A threshold can distinguish the two states \sZero and \sOne at a success rate of \SI{99.95}{\percent}.
In \Cref{sec:study:transient}, we show that this side-channel leakage can be used for a fast covert channel for Spectre attacks.

\section{Case Studies}\label{sec:casestudies}
In this section, we present three case studies based on the newly detected side channels (\cf \cref{sec:results}). 
\cref{sec:study:transient} demonstrates that the newly discovered side channels can be used for transient-execution attacks. 
They can be used in Spectre attacks to increase the space of possible gadgets, as well as in Meltdown-type attacks to increase the leakage. 
\cref{sec:study:kaslr} introduces a novel microarchitectural attack against kernel-level ASLR (KASLR) based on the results discovered by \ToolName. 
This novel KASLR break even works on the newest Intel Ice Lake and Comet Lake microarchitectures, even if all known mitigations are in place. 
\cref{sec:study:covert} shows that the \texttt{RDRAND}-based side channel can be used as a cross-core covert channel in the cloud without relying on the cache. 

\subsection{Transient-Execution Covert Channels}\label{sec:study:transient}
Transient-execution attacks~\cite{Canella2019A}, \ie Spectre- and Meltdown-type attacks, always require a microarchitectural covert channel to transfer the microarchitecturally-encoded data into the architectural state. 
Typically, these attacks rely on a cache covert channel~\cite{Canella2019A}, as also shown in the original Spectre~\cite{Kocher2019} and Meltdown~\cite{Lipp2018meltdown} paper. 
Cache-based covert channels have the advantage that they are ubiquitous, fast, and reliable~\cite{Canella2019A,Kocher2019,Lipp2018meltdown}. 
In this case study, we show that our new side channels can potentially increase the number of Spectre gadgets, and optimize the leakage for Meltdown-type attacks. 

\paragraph{Spectre Attacks.}
Bhattacharyya~\etal\cite{Bhattacharyya2019} and Schwarz~\etal\cite{Schwarz2019netspectre,Schwarz2019STL} already showed different covert channels for Spectre. 
Their covert channels are based on port contention, vector instructions, and the TLB, respectively. 
In this case study, we show that our newly discovered side channel based on \styleInstruction{AVX2} and \styleInstruction{x87-FPU} can also be used for Spectre attacks. 

We implement a proof-of-concept Spectre attack that uses this side channel as the covert channel.
Our proof of concept exploits Spectre-PHT~\cite{Kocher2019} to leak a string outside of the bounds of an array.
We can use the same gadgets as in a NetSpectre attack~\cite{Schwarz2019netspectre} and similar gadgets as used in SMoTherSpectre~\cite{Bhattacharyya2019}.
More specifically, exploiting the discovered side channels would require finding specific gadgets (conditional trigger sequence) in the victim code. 
Such gadgets could also be constructed in combination with other Spectre vulnerabilities using speculative ROP~\cite{Bhattacharyya2019,Bhattacharyya2020secrop}. %
Depending on the value of a transiently accessed bit, an \styleInstruction{AVX2} instruction is executed or not executed.
While NetSpectre simply waits for the state to be reset, we rely on the findings of \ToolName that executing an \styleInstruction{x87-FPU} instruction resets the state faster. 
The receiving end of the covert channel is again an \styleInstruction{AVX2} instruction.
We tested our code on an \styleArchitecture{Intel Core i7-9700K}, where we achieved a leakage rate of \SI{2407}{\bit/\second} with an error rate of \SI{0.43}{\percent}.
This is 2.4 times as fast as the transmission rate of the \styleInstruction{AVX}-based covert channel used in NetSpectre~\cite{Schwarz2019netspectre}.

\paragraph{Meltdown Attacks.}
In Meltdown-type attacks, both the sending and the receiving end of the covert channel are entirely attacker-controlled. 
So far, all Meltdown-type attacks~\cite{Lipp2018meltdown,Stecklina2018,Vanbulck2018foreshadow,Canella2019A,VanSchaik2019RIDL,Canella2019Fallout,Schwarz2019ZL,Ragab2021} relied on the cache and typically on \FlushReload to recover the information from the cache. 
Even though \FlushReload is extremely fast and reliable, it is still the bottleneck for leaking data~\cite{Lipp2018meltdown}. 

With \MovReload, we introduce a new cache attack for improving the leakage rate of Meltdown-type attacks. 
\MovReload is based on the discovery of \ToolName that non-temporal memory stores flush the target from the cache. 
While a cache attack that requires shared writable memory is not useful in a typical side-channel scenario, it is ideal as a fast covert channel for transient-execution attacks. 
\MovReload replaces the \texttt{CLFLUSH} instruction with a \texttt{MOVNTDQ} instruction. 
The \texttt{MOVNTDQ} instruction has a similar effect as the \texttt{CLFLUSH} instruction. It evicts the target cache line from the cache~\cite{Intel_opt}. 

\subparagraph{Reliability of Eviction.} Using L3 performance counters, we confirmed that the \texttt{MOVNTDQ} instruction indeed reliably evicts the cache line from all cache levels. 
With respect to the eviction reliability, there is no difference between \texttt{MOVNTDQ} and \texttt{CLFLUSH} or \texttt{CLFLUSHOPT}. 
Both for \MovReload and \FlushReload, we measured an F-score of 1.0 ($n$ = \SIx{10000000}). 
Furthermore, even novel cache designs~\cite{Liu2014random,Qureshi2018,Werner2019} likely do not prevent this type of eviction, as they only block the flush instruction and prevent the efficient creation of eviction sets. 

\subparagraph{Performance.} We observe one significant difference between \FlushReload and \MovReload. 
Although in both attacks, the value is evicted from all cache levels, the reload of a value flushed using \texttt{MOVNTDQ} is significantly faster on all our tested CPUs. 
On the i7-8565U, for example, reloading a value when it was flushed takes on average 253 cycles ($n$ = \SIx{20000000}) (including an \texttt{MFENCE} each before and after the memory load). 
In contrast, when the value was evicted using \texttt{MOVNTDQ}, reloading only takes 172 cycles ($n$ = \SIx{20000000}). 
Analyzing the uncore performance counters shows that this time difference for loading the data originates from the uncore (\texttt{offcore\_requests\_outstanding.cycles\_with\_data\_rd}). 
We attribute the time difference to the cache-coherency protocol.
Flushing the cache line puts the cache line into the \textit{invalid} state, while writing to the cache line puts it into the \textit{modified} state~\cite{Palanca2003clflush,Molka2015cache}. 
When loading the flushed cache line, it switches to the \textit{exclusive} state, while the \textit{modified} state stays the same. 
Due to the different behaviors of cache snooping, loading from different cache coherence states also results in different latencies~\cite{Molka2015cache}. 

\subparagraph{Results.} The faster reload time allows encoding \SIx{2.5}x more values during the transient window. 
In a Meltdown proof of concept relying on \MovReload, we can, on average, leak \SIx{7.83} bytes at once ($n$ = \SIx{100000}) (Intel i3-5010U).\footnote{We used this older CPU as the new CPUs are not affected by Meltdown.}
Previous work was only able to leak up to 3 bytes~\cite{Lipp2018meltdown,Schwarz2019ZL,Moghimi2020medusa,Ragab2021}. 

\subsection{MOVNT-based KASLR Break}\label{sec:study:kaslr}
KASLR has been subject to almost countless microarchitectural attacks in the past~\cite{Jang2016,Hund2013,Gruss2016Prefetch,Evtyushkin2016ASLR,Canella2019Fallout,Schwarz2019STL,Canella2020kaslr, Maisuradze2018speculose}.
As a response, researchers, CPU vendors, and OS maintainers have developed several countermeasures~\cite{Gens2017,Gruss2017KASLR,Canella2020kaslr,FGKASLR_v5}.
In particular, the newest 10th-generation Intel CPUs (Ice Lake and Comet Lake) are immune to many microarchitectural KASLR breaks, including the recently discovered EchoLoad attack~\cite{Canella2020kaslr}.
However, our newly-discovered side channel can be used to break KASLR even on those architectures.

Based on the discovery of \ToolName that the \texttt{MOVNT} instruction evicts a cache line, we manually evaluated whether this eviction also works for inaccessible addresses such as kernel addresses.
Previous work showed that even for Meltdown-resistant CPUs, memory loads~\cite{Vanbulck2020lvi,Canella2020kaslr} and stores~\cite{Schwarz2019STL} can infer side-channel information from the kernel. 
Although \texttt{MOVNT} could not directly evict kernel memory, we observed changes in the cache state on seemingly unrelated memory.
If the targeted kernel address is invalid, \ie not physically backed, we observe that an unrelated \texttt{MOV} on user memory issued after the \texttt{MOVNT} fails.
If the kernel address is physically backed, the \texttt{MOV} is successful. 
Hence, this allows de-randomizing the location of the kernel, effectively breaking KASLR. 

\begin{listing}
\begin{lstlisting}[style=customc]
try {
  asm volatile(
    "clflush 0(%[probe])\n"
    "movq %%rsi, (%[dummy])\n"
    "movntdqa (%[kernel]), %%xmm1\n"
    "movq (%[probe]), %%rax\n"
  ) : : [probe]"r"(probe), [dummy]"r"(dummy), 
        [kernel]"r"(kernel)
    : "rax", "xmm1", "rsi", "memory");
} catch {
  if(uncached(probe)) return MAPPED;
  else return UNMAPPED;
}
\end{lstlisting}
\caption{The main part of \KASLRBreak. 
The probe memory is uncached if the kernel address is physically backed.}
\label{lst:kaslr}
\end{listing}

\Cref{lst:kaslr} shows the minimal working example of our KASLR break, \KASLRBreak, that we created from our findings on \texttt{MOVNT}. 
A user-accessible memory address (\texttt{probe}) is flushed, followed by a write to an unrelated address, acting as a reordering barrier.
Afterward, the kernel address (\texttt{kernel}) is read using \texttt{MOVNT}.
Finally, \texttt{probe} is accessed. 
As the load from the kernel address leads to a fault, exceptions are handled using a signal handler for this code. 
After resolving the fault, the cache state of \texttt{probe} is observed, \eg using \FlushReload. 
If \texttt{probe} is cached, the kernel address is invalid, if \texttt{probe} is not cached, the kernel address is valid.

\paragraph{Root-Cause Hypothesis.}
Using performance counters, we analyzed the behavior of \KASLRBreak.
The \texttt{CLFLUSH} and load access to the same address trigger a cache-line conflict as also exploited in ZombieLoad~\cite{Schwarz2019ZL}.
Even though, at first, the write to \texttt{dummy} seems unrelated, it is guaranteed to be ordered with \texttt{CLFLUSH}~\cite{Intel_vol2} and hence influences the overall timing of the executed code in the processor pipeline.
Alternatively, this line can also be removed entirely (depending on the CPU) or replaced by a different method to add a delay, \eg using a dummy loop.
However, adding a serializing instruction, such as a fence, breaks the attack, as it forces the \texttt{CLFLUSH} to retire, preventing the cache-line conflict with the load. 
If \texttt{kernel} is physically backed, we observe a page-table walk (\texttt{dtlb\_load\_misses.miss\_causes\_a\_walk}).
If \texttt{kernel} is \emph{not} physically backed, we observe 2 page-table walks, \ie the page-table walk is repeated. 
That is in agreement with Canella~\etal\cite{Canella2020kaslr}, showing that loads from non-present kernel pages are re-issued. 
As this case takes longer~\cite{Jang2016} and faults are only detected at the retirement of instructions, it gives other out-of-order executed instructions more time to execute. 
We hypothesize that if the kernel address is unmapped, the processor has a long-enough speculation window to execute the flush, write, and the last load.
As a result of this, the last load brings \texttt{probe} back to the cache.
In the case of a mapped kernel address, the processor detects the fault earlier and hence stops the execution before the last load was issued.
As a result, \texttt{probe} is cached if \texttt{kernel} is not physically backed, and not cached if \texttt{kernel} is physically backed. 
The observed performance counters back this hypothesis.
For an unmapped address, \texttt{mem\_load\_retired\_l3\_miss} shows fewer events. 
However, the number of cycles spent waiting for memory (\texttt{cycle\_activity.cycles\_l3\_miss}) is slightly higher.
This indicates that there are ongoing load instructions that never retire, backing the hypothesis that the last load is only executed transiently when the address is unmapped.

\begin{table}
\caption{The evaluated CPUs for the KASLR break.}
\label{tab:kaslr-cpus}
\adjustbox{max width=\hsize}{
 \begin{tabular}{lrrr}
 \toprule
 \textbf{CPU (Microarchitecture)}       & \textbf{Accuracy (idle)} & \textbf{Accuracy (stress)} & \textbf{Runtime} \\
  \midrule                                                                                    
  Intel Core i5-3230M (Ivy Bridge)      & \SI{99}{\percent}                     & \SI{97}{\percent}                       & \SI{34}{\milli\second} \\
  Intel Core i5-4690 (Haswell)          & \SI{100}{\percent}                     & \SI{99}{\percent}                      & \SI{221}{\milli\second} \\
  Intel Core i3-5010U (Broadwell)       & \SI{99}{\percent}                     & \SI{97}{\percent}                       & \SI{5}{\milli\second} \\    %
  Intel Core i7-6700K (Skylake)         & \SI{99}{\percent}                     & \SI{98}{\percent}                & \SI{9}{\milli\second} \\
  Intel Core i7-8565U (Whiskey Lake)    & \SI{100}{\percent}                     & \SI{92}{\percent} & \SI{6}{\milli\second} \\
  Intel Core i7-9700K (Coffee Lake)     & \SI{100}{\percent}                    & \SI{98}{\percent}                       & \SI{102}{\milli\second} \\ %
  Intel Core i9-9980HK (Coffee Lake)    & \SI{99}{\percent}                     & \SI{99}{\percent}                       & \SI{65}{\milli\second} \\  %
  Intel Core i3-1005G1 (Ice Lake)       & \SI{96}{\percent}                     & \SI{96}{\percent}                       & \SI{300}{\milli\second} \\ 
  Intel Core i7-10510U (Comet Lake)     & \SI{99}{\percent}                     & \SI{97}{\percent}                       & \SI{84}{\milli\second} \\
  Intel Celeron J4005 (Gemini Lake)     & \SI{99}{\percent}                     & \SI{99}{\percent}                       & \SI{349}{\milli\second} \\
  Intel Xeon Platinum 8124M (Skylake-SP)      & \SI{99}{\percent}                     & \SI{99}{\percent}                        & \SI{318}{\milli\second} \\
  \bottomrule
 \end{tabular}
}
\end{table}

\paragraph{Applicability.}
We tested our microarchitectural KASLR break on Intel CPUs from the $3^{rd}$ to the $10^{th}$ generation, \ie from Ivy Bridge to Comet Lake. 
As shown in \cref{tab:kaslr-cpus}, we used desktop (Core), server (Xeon), and mobile (Celeron) CPUs. 

In contrast, we experimentally verified that EchoLoad~\cite{Canella2020kaslr}, which works on a large range of Intel CPUs from 2010 to 2019, does not work on Ice Lake or Comet Lake. 
We confirm that the KASLR break is operating-system agnostic by successfully mounting it on Linux and Windows 10.

In the case of KPTI, \ie on CPUs that are not Meltdown-resistant, the KASLR break detects the trampoline used to switch to the kernel. 
Otherwise, if the CPU is Meltdown-resistant or KPTI is disabled, the KASLR break detects the start of the kernel image. 
As an unprivileged attacker can read out the state of KPTI and whether the CPU is vulnerable to Meltdown, the attacker always knows the start of the kernel image. 
Moreover, as the kernel image itself is not randomized, knowing the kernel version and the start of the kernel image is sufficient to calculate the location of any kernel part. 

Additionally, we tested the KASLR break by simulating a realistic environment by artificially raising the pressure on the CPU and memory subsystem using the \styleCommand{stress} utility.
We still observe success rates ranging from 92\% to 99\% for the different microarchitectures ($n$ = \SIx{100}). 
Furthermore, we verified the KASLR break in a cloud scenario by testing it on an \styleArchitecture{Intel Xeon Platinum 8124M} in the AWS cloud.

\paragraph{Performance.} 
On average, our KASLR break detects the start of the kernel image within \SI{136}{\milli\second} ($n$ = \SIx{1100}) 
While not the fastest microarchitectural KASLR break, it is on par with other microarchitectural KASLR breaks~\cite{Canella2020kaslr}.

\subsection{\styleInstruction{RDRAND} Covert Channel in the Cloud}\label{sec:study:covert}
\ToolName discovered a timing leakage in the \texttt{RDRAND} instruction on both Intel and AMD CPUs.
In this section, we present a cross-core covert channel based on these timing differences. 
We evaluate the capacity in a cross-thread scenario (\Cref{sec:rdrand-samecore}), and across cores and VMs (\Cref{sec:rdrand-acrosscores}). 
Finally, we analyze the leakage reason (\Cref{sec:rdrand-reason}).

\begin{table}
\caption{The evaluated CPUs for the \texttt{RDRAND} covert channel.}
\label{tab:rdrand-cpus}
\centering
\adjustbox{max width=\hsize}{
 \begin{tabular}{llrrrr}
 \toprule
  \multirow{2}{*}{\textbf{CPU}} & \multirow{2}{*}{\textbf{Setup}} & \multicolumn{2}{c}{Cross-HT} & \multicolumn{2}{c}{Cross-Core} \\
  & & \textbf{Speed} & \textbf{Error} & \textbf{Speed} & \textbf{Error} \\
  \midrule
  Intel Core i5-3230M & Lab & \SI{133.3}{\bit/\second}                     & \SI{8.87}{\percent}      & \SI{133.3}{\bit/\second}                     & \SI{0.05}{\percent}       \\
  Intel Core i3-5010U & Lab & \SI{666.7}{\bit/\second} & \SI{0.30}{\percent} & \SI{333.3}{\bit/\second} & \SI{1.82}{\percent} \\  %
  Intel Core i7-8565U & Lab & \SI{400.0}{\bit/\second} & \SI{0.65}{\percent} & \SI{166.7}{\bit/\second} & \SI{0.63}{\percent} \\
  Intel Core i9-9980HK & Lab & \SI{500.0}{\bit/\second} & \SI{0.76}{\percent} & \SI{117.6}{\bit/\second} & \SI{9.25}{\percent} \\ %
  Intel Core i3-1005G1 & Lab & \SI{1000.0}{\bit/\second} & \SI{0.37}{\percent} & \SI{1000.0}{\bit/\second} & \SI{0.00}{\percent} \\ %
  Intel Xeon E5-2686 v4 & Cloud & \SI{500.0}{\bit/\second} & \SI{0.21}{\percent} & \SI{333.3}{\bit/\second} & \SI{2.48}{\percent} \\
  Intel Xeon E5-2666 v3 & Cloud & \SI{666.7}{\bit/\second} & \SI{2.64}{\percent} & \SI{95.2}{\bit/\second} & \SI{0.88}{\percent} \\
  AMD Ryzen 5 2500U & Lab & \SI{48.8}{\bit/\second} & \SI{2.80}{\percent} & \SI{48.8}{\bit/\second} & \SI{2.00}{\percent} \\
  AMD Ryzen 5 3550H & Lab & \SI{666.7}{\bit/\second} & \SI{2.10}{\percent} & \SI{500.0}{\bit/\second} & \SI{2.50}{\percent} \\
  
  \bottomrule
 \end{tabular}
}
\end{table}

\subsubsection{Setup}
The setup consists of a sender and a receiver application. 
In our proof-of-concept implementation, sender and receiver are simply time-synchronized, \ie they rely on a common time source such as the timestamp counter. 
To send a `1'-bit, the sender repeatedly executes the \texttt{RDRAND} instruction for a fixed time $\tau$. 
To send a `0'-bit, the sender idles for $\tau$. 
The receiver measures the latency of the \texttt{RDRAND} instruction over a period of $\tau$. 
The latency directly corresponds to the sent bit, \ie a high latency is caused by a `1'-bit, and a low latency is caused by a `0'-bit. 
We note that this setup is not optimal, as there are more advanced techniques for synchronization, including error correction~\cite{Wu2012,Evtyushkin2016RNG,Maurice2017Hello}.
However, our goal is to show the feasibility and the noise-resistance of this channel, not how far it can be optimized using better engineering. 

\subsubsection{Same-core Leakage}
\label{sec:rdrand-samecore}
We evaluated an \texttt{RDRAND}-based covert channel across hyperthreads to estimate the maximum capacity of this channel. 
Note that the leakage in a cross-hyperthread channel is boosted by port contention as well~\cite{Aldaya2018,Bhattacharyya2019}. 
Moreover, on Intel CPUs, Intel documents that the microcode update preventing SRBDS~\cite{Ragab2021} serializes RDRAND executions on the same core~\cite{Intel2020srbds}. 
Hence, to rule out any influence of the microcode fixes, we evaluated the channel with and without the active patches. 
As AMD CPUs are not susceptible to SRBDS, there is no microcode influence to rule out. 
As \Cref{tab:rdrand-cpus} shows, we verified the covert channel on all Intel microarchitectures since at least the Ivy Bridge microarchitecture, and also on the AMD Zen and Zen+ microarchitecture.
We achieve the best results on the newest microarchitectures, with \SI{1000}{\bit/\second} (\SI{0}{\percent} error) on Intel and \SI{666.7}{\bit/\second} (\SI{2.1}{\percent} error) on AMD.
While a same-core channel is usually irrelevant, it shows the upper bound of the leakage achievable across cores. 

\subsubsection{Cross-core Leakage}
\label{sec:rdrand-acrosscores}
In addition to the expected leakage across hyperthreads, we evaluate the channel across physical cores. 

\paragraph{Local Environment.}
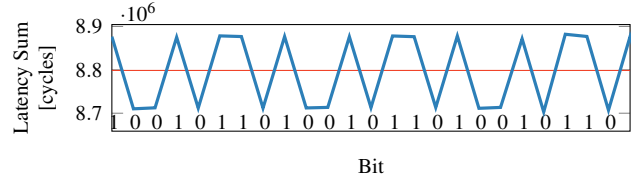
\begin{figure}
 \input{images/rdrand-ice-lake.tikz}
 \caption{Using the \texttt{RDRAND} covert channel to send the bit stream \texttt{100101101001011010010110...} from one CPU core to a different physical core (Intel Core i3-1005G1).}
 \label{fig:rdrand-local}
\end{figure}
\Cref{fig:rdrand-local} shows a cross-core transmission in a local environment. 
While the signal is weaker than in the cross-hyperthread scenario, we still manage to transmit data reliably. 
As shown in \Cref{tab:rdrand-cpus}, the channel achieves up to \SI{1000}{\bit/\second} with a low error rate down to \SI{0}{\percent}. 

\paragraph{AWS Cloud.}
To further evaluate the applicability of the covert channel in a real-world scenario, we mounted it between two virtual machines running in the AWS cloud.
To ensure that we do not interfere with other users, we used a dedicated C3 host with an Intel Xeon E5-2666 v3.
We were able to transmit \SI{95.2}{\bit/\second} across two different virtual machines running on the same CPU with an error rate of \SI{0.88}{\percent}.
Additionally, the host had a third virtual machine running to simulate realistic noise. 
For completeness, we also verified that the covert channel works across hyperthreads and cores inside a single virtual machine in this setup (\cf \Cref{tab:rdrand-cpus}).

\paragraph{Comparison to Other Cross-Core Covert Channels.}
\begin{table}
\caption{Transmission and error rates of state-of-the-art cross-core covert channels sorted by transmission speed.}
\label{tab:covert-speed}
\adjustbox{max width=\hsize}{
 \begin{tabular}{lrr}
  \toprule
  \textbf{Covert channel (Element)} & \textbf{Speed} & \textbf{Error rate} \\
  \midrule
  Liu~\etal\cite{Liu2015Last} (L3) & \SI{600}{\kilo\bit/\second} & \SI{1.00}{\percent} \\
  Pessl~\etal\cite{Pessl2016} (DRAM) & \SI{411}{\kilo\bit/\second} & \SI{4.11}{\percent} \\
  Maurice~\etal\cite{Maurice2017Hello} (L3) & \SI{362}{\kilo\bit/\second} & \SI{0.00}{\percent} \\
  Evtyushkin~\etal\cite{Evtyushkin2016RNG} (RDSEED) & \SI{71}{\kilo\bit/\second} & \SI{0.00}{\percent} \\ %
  Ragab~\etal\cite{Ragab2021} (CPUID) & \SI{24}{\kilo\bit/\second} & \SI{5.00}{\percent} \\
  \textbf{Ours (RDRAND)} & \SI{1000}{\bit/\second} & \SI{0.00}{\percent} \\
  Maurice~\etal\cite{Maurice2015C5} (L3) & \SI{751}{\bit/\second} & \SI{5.70}{\percent} \\
  Wu~\etal\cite{Wu2012} (memory bus) & \SI{747}{\bit/\second} & \SI{0.09}{\percent} \\
  Semal~\etal\cite{Semal2020covert} (memory bus) & \SI{480}{\bit/\second} & \SI{5.46}{\percent} \\
  Schwarz~\etal\cite{Schwarz2017Timers} (DRAM) & \SI{11}{\bit/\second} & \SI{0.00}{\percent} \\
  \bottomrule
 \end{tabular}
}
\end{table}
\Cref{tab:covert-speed} shows a comparison of the transmission speed for state-of-the-art cross-core covert channels. 
While the \texttt{RDRAND}-based covert channel is much slower than modern cache-based covert channels, it has two huge advantages. 
First, there are no performance counters for the hardware random number generator. 
Thus, this channel cannot be easily detected or prevented by current approaches relying on performance counters~\cite{Chiappetta2015,Payer2016,Irazoqui2018mascat,Herath2015}. 
We also used the open-source HexPADS framework~\cite{Payer2016} to verify that it cannot detect the covert channel.
Second, in contrast to memory-based covert channels, this channel is agnostic to any typical system noise caused by memory accesses on the sender core. 
As typical workloads do not execute \texttt{RDRAND} in a high frequency, we do not see a high impact on the transmission rate, even for high workloads.
We verified that by running the Linux tool \texttt{stress} for both the CPU and the memory on the sender core does not prevent the covert channel. 
Even in this scenario, with an extremely high load of \SI{100}{\percent} on the sibling hyperthread, we manage to transmit \SI{500.0}{\bit/\second} with an error rate of \SI{7.34}{\percent}.

Furthermore, as our covert channel does not rely on the memory subsystem, defenses proposed against cache attacks~\cite{Wang2007,Zhang2011,Zhang2013,Liu2014random,Qureshi2018,Werner2019} do not prevent our channel. 
Even existing partitioning features, such as Intel CAT, which can be used to prevent cache-based cross-VM covert channels~\cite{Liu2016catalyst} do not affect the \texttt{RDRAND}-based covert channel.

\subsubsection{Explanation for RDRAND Side Channel}
\label{sec:rdrand-reason}
As the hardware random number generator is shared across all cores, simultaneous use by multiple cores leads to contention. 
Hence, as with many cross-core covert channels~\cite{Wu2012,Maurice2015C5,Pessl2016,Evtyushkin2016RNG}, the root cause is the contention of a resource shared across cores, such as the L3 cache or the memory bus. 
However, in contrast to previous covert channels, we could not identify any performance counters related to \texttt{RDRAND}. 
While this makes the analysis more difficult, it also increases the stealthiness of the channel, as it cannot be detected easily. 

While previous work showed that the \texttt{RDSEED} instruction can exhaust the hardware random-number generator (RNG)~\cite{Evtyushkin2016RNG}, the \texttt{RDRAND} instruction has not been analyzed for side-channel leakage. 
Moreover, Evtyushkin~\etal\cite{Evtyushkin2016RNG} only exploited an architectural value, \ie a cleared carry flag, indicating that the RNG is exhausted, and not differences in the execution time. 
At first glance, it might seem obvious that \texttt{RDRAND} also suffers from exhaustion as it fundamentally relies on the \texttt{RDSEED} instruction. 
\texttt{RDSEED} is quickly exhausted, as it provides the randomness directly from the hardware element. 
However, Evtyushkin~\etal\cite{Evtyushkin2016RNG} observed that \texttt{RDRAND} provides the numbers from a pseudo-RNG and can thus provide continuous streams of numbers. 
We confirm that the \texttt{RDRAND}-based leakage is \emph{not} due to exhaustion. 
While measuring the timing differences, the instruction does not indicate that the RNG is exhausted, \ie the carry flag was always set~\cite{Intel_vol3}. 

We additionally ruled out the microcode updates preventing CrossTalk~\cite{Ragab2021} as a cause for the timing differences. 
While these updates reduce the bandwidth of \texttt{RDRAND} across hyperthreads due to serialization, they do not affect the cross-core behavior~\cite{Intel2020srbds}. 
We verified that by successfully mounting the covert channel with and without the microcode update, and also by disabling the mitigation on patched systems via the \texttt{IA32\_MCU\_OPT\_CTRL} model-specific register.

\section{Discussion}\label{sec:discussion}

With \ToolName, we present a generic approach for detecting timing-based side channels. 
Our current prototype still has several limitations preventing it from finding even more side channels. 
However, these are not conceptual limitations. 
It would merely require a lot more engineering to solve them. 
In the current version, we only consider side channels where the timing difference is around 100 cycles. 
Any side channel with a smaller timing difference, \eg \FlushFlush~\cite{Gruss2016Flush}, CacheBleed~\cite{Yarom2017cachebleed} or the AMD way predictor~\cite{Lipp2020takeaway}, is currently not reported. 
One practical reason is that \ToolName runs on a commodity Linux system, where it is tough to eliminate all influences on the measurement. 
Even when isolating cores, several microarchitectural elements are shared across all cores, there are still remaining interrupts, and the power management of the CPU can change the CPU frequency, \eg for thermal reasons. 
Hence, to reliably detect small timing differences, \ToolName would have to run on a custom operating system designed for microarchitectural research, such as Sushi Roll~\cite{Falk2019sushi}. 
In line with related work~\cite{Fogh2016shotgun,Gras2020}, our prototype only considers sequences consisting of one instruction.
As a consequence, eviction-based side channels such as \EvictReload, \EvictTime, \PrimeProbe, or Reload+Refresh are not detected. 
However, related work~\cite{Gruss2016Row,Vila2019,Vila2020cache} showed that eviction strategies can also be found automatically. 
Moreover, for specific problems, the search space can be reduced by mutating existing instruction sequences (similar to Medusa~\cite{Moghimi2020medusa}) or instruction operands instead of randomly generating them. 
Therefore, \ToolName can be augmented by these techniques to also find eviction-based side channels and support multi-instruction sequences (\eg fault suppression). 
Furthermore, using performance counters, power (RAPL), and debug interfaces (Intel VISA/ITP-XDP) as feedback mechanisms, the fuzzer could monitor resource usage and microarchitectural conflicts to guide the sequence generation process. 
This would allow finding eviction-based channels: (i) Start with multiple loads as a reset sequence, (ii) Mutate the loaded addresses while maximizing (guidance) the cache miss count until a time difference is detected.

Still, despite these current limitations of the prototype, \ToolName discovered novel timing-based side channels within hours of runtime. 
These side channels led to the discovery of a new microarchitectural KASLR break, a previously unknown cross-VM covert channel, and an improvement for transient-execution attacks. 
Hence, we argue that \ToolName is a useful tool for automating the search for timing-based side channels that can also be used by CPU vendors to detect such side channels introduced by new ISA extensions automatically. 

Also, \ToolName can be extended to other architectures, \eg ARMv8, with relative ease.  
To this end, the main parts that need to be adapted are the code generation stage, particularly the offline phase to construct possible instruction variants, and the execution stage. 
The current implementation of \ToolName uses inlined instructions to measure the execution time, which would need to be changed for the target architecture (see Section~\ref{sec:implementation}). 
However, this task can be simplified by refining the current approach to use other timing primitives~\cite{Lipp2016}.

\section{Conclusion}\label{sec:conclusion}
Our findings illustrate that prior side channels targeted only a subset of many micro-architectural changes.
We show several additional, undocumented instruction side effects that attackers can leverage for security-critical side channels.
This has severe implications to existing and future side-channel defenses, as each of them is based on a specific threat model that frames (known) attack capabilities.
We, therefore, see our proposed fuzzing-based technique as the first \emph{systematic}, \emph{generic}, and \emph{automated} attempt to fast-forward the arms race of detecting (and then, ultimately, defending against) such side channels.
The newly discovered side channels and their application to three use cases raise our confidence that \ToolName can indeed support this endeavor.
When used during the CPU design stage, \ToolName helps to eliminate---or at least to document---side channels early on.
For this reason, we released \ToolName as an open-source tool.

\ifAnon

\else
\section*{Acknowledgments}
We thank the anonymous reviewers and our shepherd, Mathias Payer, for their helpful comments and suggestions that substantially helped in improving the paper, as well as Moritz Lipp (Graz University of Technology) for feedback on an earlier version of the paper. 
Furthermore, we thank the Saarbrücken Graduate School of Computer Science for their funding and support for Daniel Weber.
This work partially was supported by grant from the German Federal
Ministry of Education and Research (BMBF) through funding for the CISPA-Stanford Center for Cybersecurity (FKZ:13N1S0762).
\fi

\bibliographystyle{plainurl}
\bibliography{main}

\appendix
\FloatBarrier

\section{Clustering Results}\label{sec:appendix:cluster}

\begin{table}
\centering
  \caption{Cluster Results For Intel Microarchitectures.}
  \label{tbl:clusters}
  \adjustbox{max width=\hsize}{
  \begin{tabular}{lrrr}
  \toprule
\textbf{CPU Name} & \textbf{Found} & \textbf{Extension} & \textbf{\MeSeq-\TrSeq only} \\
\midrule
Intel Core i7-9750H  & \SIx{68597} & \SI{186}{clusters}  & \SI{16}{clusters} \\ %
Intel Core i5-4690   & \SIx{51468} & \SI{168}{clusters}  & \SI{19}{clusters} \\ %
Intel Core i7-9700K  & \SIx{27512} & \SI{104}{clusters}  & \SI{26}{clusters} \\ %
\bottomrule
  \end{tabular}
  }
\end{table}

\Cref{tbl:clusters} shows the clustering results for the CPUs on which \ToolName ran. 
\ToolName found multiple thousand side channels that were clustered based on the instruction extension of \TrSeq, \MeSeq, and \ReSeq, resulting in 100 to 200 clusters. 
However, as \ReSeq is typically not involved in the actual leakage, clustering based on the instruction extension of only \TrSeq and \MeSeq results in a smaller number of clusters. 

\end{document}

%% file: images/states.tikz
\begin{tikzpicture}[transform shape,scale=0.8]

\draw[draw=black,fill=black!5] (0,0) ellipse (0.75cm and 0.5cm) node[midway] {S0};

\draw[draw=black,fill=black!5] (3,0) ellipse (0.75cm and 0.5cm) node[midway,xshift=3cm] {S1};

\draw[out=90,in=90,->,>=stealth,looseness=4] (-0.2,0.5) to (0.2,0.5) node[midway,above,yshift=1cm] {\small Reset Seq.};

\draw[out=90,in=90,->,>=stealth,looseness=4] (2.8,0.5) to (3.2,0.5) node[midway,above,yshift=1cm,xshift=3cm] {\small Trigger Seq.};

\draw[out=30,in=150,->,>=stealth] (0.65,0.25) to (2.35,0.25) node[above,midway,xshift=1.5cm,yshift=0.4cm] {\small Trigger Seq.};

\draw[out=210,in=-30,->,>=stealth] (2.35,-0.25) to (0.65,-0.25) node[above,midway,xshift=1.5cm,yshift=-1cm] {\small Reset Seq.};

\end{tikzpicture}

%% file: images/overview.tikz
\resizebox{\hsize}{!}{
\centering
\begin{tikzpicture}[yscale=0.75]
\draw[draw=none,fill=black!5] (-3,-1) rectangle +(3,2) node[above,xshift=-1cm] {\small Offline};
\draw[densely dotted] (0,-1) rectangle +(3,2) node[above,xshift=-1.5cm] {\small \circled{1} Generation};
\draw[densely dotted] (3,-1) rectangle +(2,2) node[above,xshift=-1cm] {\small \circled{2} Execution};
\draw[densely dotted] (5,-1) rectangle +(4,2) node[above,xshift=-2cm] {\small \circled{3} Confirmation};
\draw[densely dotted] (9,-1) rectangle +(2,2) node[above,xshift=-1cm] {\small \circled{4} Clustering};

\node at (-2,0) {\parbox{2cm}{\centering \tiny \includegraphics[width=1cm]{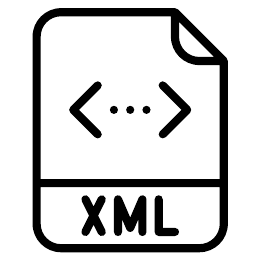}\\ISA}};
\node at (0,0) {\parbox{2cm}{\centering \tiny \includegraphics[width=1cm]{images/xml}\\Instructions}};
\node at (2,0) {\parbox{2cm}{\centering \tiny \includegraphics[width=1cm]{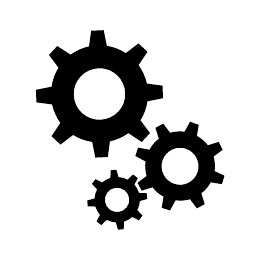}\\Triple Generation}};
\node at (8,0) {\parbox{2cm}{\centering \tiny \includegraphics[width=1cm]{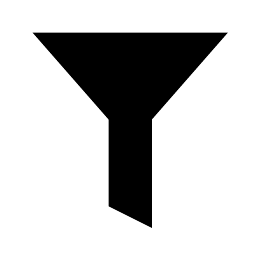}\\Leaking Triples}};
\node at (6,0) {\parbox{2cm}{\centering \tiny \includegraphics[width=1cm]{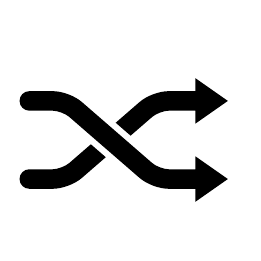}\\Randomized Execution}};
\node at (10,0) {\parbox{2cm}{\centering \tiny \includegraphics[width=1cm]{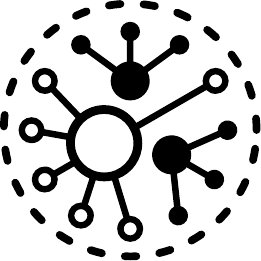}\\Clustering}};
\node at (12,0) {\parbox{2cm}{\centering \tiny \includegraphics[width=1cm]{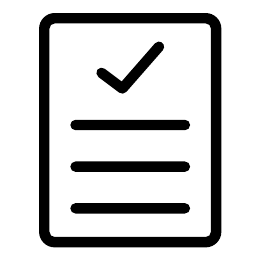}\\Report}};
\node at (4,0) {\parbox{2cm}{\centering \tiny \includegraphics[width=1cm]{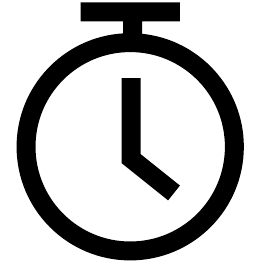}\\Timing Measurement}};

\draw[->,>=stealth,thick,out=0,in=180] (-1.25,0) to (-0.5,0);
\draw[->,>=stealth,thick,out=0,in=180] (0.75,0) to (1.5,0);
\draw[->,>=stealth,thick,out=0,in=180] (2.75,0) to (3.25,0);
\draw[->,>=stealth,thick,out=0,in=180] (4.75,0) to (5.25,0);
\draw[->,>=stealth,thick,out=0,in=180] (6.75,0) to (7.5,0);
\draw[->,>=stealth,thick,out=0,in=180] (8.5,0) to (9.25,0);
\draw[->,>=stealth,thick,out=0,in=180] (10.75,0) to (11.5,0);

\end{tikzpicture}
}

%% file: images/measure.tikz
\resizebox{\hsize}{!}{
\begin{tikzpicture}[yscale=0.75]

\draw[rounded corners,thick,dotted,draw=red!60!black,fill=red!10] (1, -1.25) rectangle +(5.5, 2.65) {};
\draw[rounded corners,thick,dashed,draw=blue!40!black,fill=blue!5] (1, 1.75) rectangle +(3, 2.65) {};

\draw (-0.5,1.55) -- (6.5,1.55);

\node at (0,0) {\parbox{1.5cm}{\centering\includegraphics[width=1.25cm]{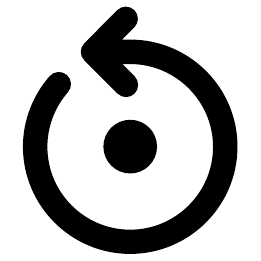}\\$Seq_{reset}$}};
\node at (2.5,0) {\parbox{1.5cm}{\centering\includegraphics[width=1.25cm]{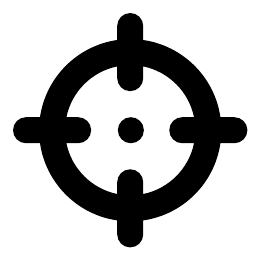}\\$Seq_{trigger}$}};
\node at (5,0) {\parbox{1.5cm}{\centering\includegraphics[width=1.25cm]{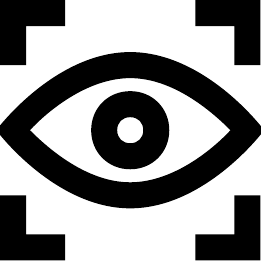}\\$Seq_{measure}$}};

\node at (0,3) {\parbox{1.5cm}{\centering\includegraphics[width=1.25cm]{images/reset}\\$Seq_{reset}$}};
\node at (2.5,3) {\parbox{1.5cm}{\centering\includegraphics[width=1.25cm]{images/view}\\$Seq_{measure}$}};

\node at (7.75,3) {\parbox{2.5cm}{\centering\includegraphics[width=1cm]{images/timing}\\\textcolor{blue!30!black}{Cold path} $S_0$}};

\node at (7.75,0) {\parbox{2.5cm}{\centering\includegraphics[width=1cm]{images/timing}\\\textcolor{red!40!black}{Hot path} $S_1$}};

\node at (7.75,1.5) {\parbox{1.5cm}{\centering\includegraphics[width=0.75cm]{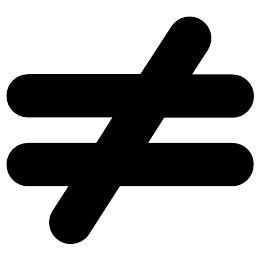}}};
\node at (10,1.5) {\parbox{1.5cm}{\centering\includegraphics[width=1.5cm]{images/report}}};
\draw[->,>=stealth,ultra thick] (8.5,1.5) to (9.25,1.5);

\end{tikzpicture}
}

%% file: images/rand.tikz
\begin{tikzpicture}
\begin{axis}[
style={font=\footnotesize},
scaled y ticks=false,
width=\hsize,
height=3.2cm,
xmin=850,
xmax=1250,
legend style={at={(0.65,1)}},
xlabel={Execution time [cycles]},
ylabel={Observations},
xtick={900,1050,...,1200},
ytick={0,7000,14000}
]
\addplot+[blue,thick,mark=none] table[x=cycles,y=with,col sep=comma] {data/rand.csv};
\addplot+[red,thick,mark=none,densely dashed] table[x=cycles,y=without,col sep=comma] {data/rand.csv};

\end{axis}
\end{tikzpicture}

%% file: images/xsave.tikz
\begin{tikzpicture}
\begin{axis}[
style={font=\footnotesize},
scaled y ticks=false,
width=\hsize,
height=3.2cm,
xmin=70,
xmax=280,
legend style={at={(0.65,1)}},
xlabel={Execution time [cycles]},
ylabel={Observations},
xtick={80,160,240}
]
\addplot+[blue,thick,mark=none] table[x=cycles,y=with,col sep=comma] {data/xsave.csv};
\addplot+[red,thick,mark=none,densely dashed] table[x=cycles,y=without,col sep=comma] {data/xsave.csv};

\end{axis}
\end{tikzpicture}

%% file: images/mmx.tikz
\begin{tikzpicture}
\begin{axis}[
style={font=\footnotesize},
scaled y ticks=false,
width=\hsize,
height=3.2cm,
xmin=25,
xmax=130,
legend style={at={(0.65,1)}},
xlabel={Execution time [cycles]},
ylabel={Observations}
]
\addplot+[blue,thick,mark=none] table[x=cycles,y=with,col sep=comma] {data/mmx.csv};
\addplot+[red,thick,mark=none,densely dashed] table[x=cycles,y=without,col sep=comma] {data/mmx.csv};

\end{axis}
\end{tikzpicture}

%% file: images/avx-pause.tikz
\begin{tikzpicture}
\begin{axis}[
style={font=\footnotesize},
scaled y ticks=false,
width=\hsize,
height=3.2cm,
xmin=15,
xmax=300,
legend style={at={(0.65,1)}},
xlabel={Execution time [cycles]},
ylabel={Observations},
ytick={0,40000,80000}
]
\addplot+[blue,thick,mark=none] table[x=cycles,y=with,col sep=comma] {data/avx-pause.csv};
\addplot+[red,thick,mark=none,densely dashed] table[x=cycles,y=without,col sep=comma] {data/avx-pause.csv};

\end{axis}
\end{tikzpicture}

%% file: images/avx-x87.tikz
\begin{tikzpicture}
\begin{axis}[
style={font=\footnotesize},
scaled y ticks=false,
width=\hsize,
height=3.2cm,
xmin=10,
xmax=220,
legend style={at={(0.65,1)}},
xlabel={Execution time [cycles]},
ylabel={Observations},
xtick={20,100,180}
]
\addplot+[blue,thick,mark=none] table[x=cycles,y=with,col sep=comma] {data/avx-x87.csv};
\addplot+[red,thick,mark=none,densely dashed] table[x=cycles,y=without,col sep=comma] {data/avx-x87.csv};

\end{axis}
\end{tikzpicture}

%% file: images/rdrand-ice-lake.tikz
\begin{tikzpicture}
\begin{axis}[
style={font=\footnotesize},
xlabel={Bit},
ylabel={\parbox{2cm}{\centering Latency Sum [cycles]}},
x tick style={draw=none},
xticklabels={,,},
width=\hsize,
xmin=0,
ymin=8658602,
xmax=24,
height=3cm,
]

\draw[red] (axis cs: 0,8798602) rectangle (axis cs: 24,8798602);
\node at (axis cs: 0.1,8680000) {1};
\node at (axis cs: 1.1,8680000) {0};
\node at (axis cs: 2.1,8680000) {0};
\node at (axis cs: 3.1,8680000) {1};
\node at (axis cs: 4.1,8680000) {0};
\node at (axis cs: 5.1,8680000) {1};
\node at (axis cs: 6.1,8680000) {1};
\node at (axis cs: 7.1,8680000) {0};
\node at (axis cs: 8.1,8680000) {1};
\node at (axis cs: 9.1,8680000) {0};
\node at (axis cs: 10.1,8680000) {0};
\node at (axis cs: 11.1,8680000) {1};
\node at (axis cs: 12.1,8680000) {0};
\node at (axis cs: 13.1,8680000) {1};
\node at (axis cs: 14.1,8680000) {1};
\node at (axis cs: 15.1,8680000) {0};
\node at (axis cs: 16.1,8680000) {1};
\node at (axis cs: 17.1,8680000) {0};
\node at (axis cs: 18.1,8680000) {0};
\node at (axis cs: 19.1,8680000) {1};
\node at (axis cs: 20.1,8680000) {0};
\node at (axis cs: 21.1,8680000) {1};
\node at (axis cs: 22.1,8680000) {1};
\node at (axis cs: 23.1,8680000) {0};

\addplot+[very thick,no marks] table[x=time,y=latency,col sep=comma] {data/rdrand-ice-lake.csv};
\end{axis}
\end{tikzpicture}